\documentclass[aps,prc,twocolumn,showpacs,superscriptaddress]{revtex4-1}

\usepackage[T1]{fontenc}
\usepackage[utf8]{inputenc}
\usepackage[english]{babel}

\usepackage{graphicx}
\usepackage{tabularx}
\usepackage{booktabs}
\usepackage{amssymb}
\usepackage{xspace}
\usepackage{caption,subcaption}
\usepackage{ragged2e}
\DeclareCaptionJustification{justified}{\justifying}
\newcolumntype{R}{>{\raggedleft\arraybackslash}X}%
\newcolumntype{C}{>{\centering\arraybackslash}X}%
\newcolumntype{L}{>{\raggedright\arraybackslash}X}%

\usepackage[version=3]{mhchem}

\newcommand{\hades}{HADES\xspace}

\newcommand{\eq}{\begin{equation}}
\newcommand{\eeq}{\end{equation}}

\usepackage{rotating}



\usepackage{xfrac}
\usepackage[olditem,oldenum]{paralist}

\usepackage{hyperref}
\hypersetup{
  plainpages=false,
  unicode=true,
  bookmarksopen=true,
  linktocpage=true,
  pdfauthor={Witold Przygoda},
}

\begin{document}

\title{$\Delta(1232)$ Dalitz decay in proton-proton collisions at T=1.25 GeV\\measured with HADES}

\newcommand{\AFFILAA}{Institute of Physics, Slovak Academy of Sciences,
84228~Bratislava, Slovakia}
\newcommand{\AFFILAB}{LIP-Laborat\'{o}rio de Instrumenta\c{c}\~{a}o e
F\'{\i}sica Experimental de Part\'{\i}culas , 3004-516~Coimbra, Portugal}
\newcommand{\AFFILAC}{Smoluchowski Institute of Physics, Jagiellonian
University of Cracow, 30-059~Krak\'{o}w, Poland}
\newcommand{\AFFILAD}{GSI Helmholtzzentrum f\"{u}r Schwerionenforschung GmbH,
64291~Darmstadt, Germany}
\newcommand{\AFFILAE}{Technische Universit\"{a}t Darmstadt, 64289~Darmstadt,
Germany}
\newcommand{\AFFILAF}{Institut f\"{u}r Strahlenphysik, Helmholtz-Zentrum
Dresden-Rossendorf, 01314~Dresden, Germany}
\newcommand{\AFFILAG}{Joint Institute for Nuclear Research, 141980~Dubna,
Russia}
\newcommand{\AFFILAH}{Institut f\"{u}r Kernphysik, Goethe-Universit\"{a}t,
60438 ~Frankfurt, Germany}
\newcommand{\AFFILAI}{Excellence Cluster 'Origin and Structure of the
Universe', 85748~Garching, Germany}
\newcommand{\AFFILBJ}{Physik Department E12, Technische Universit\"{a}t
M\"{u}nchen, 85748~Garching, Germany}
\newcommand{\AFFILBA}{II.Physikalisches Institut, Justus Liebig Universit\"{a}t
Giessen, 35392~Giessen, Germany}
\newcommand{\AFFILBB}{Institute for Nuclear Research, Russian Academy of
Sciences, 117312~Moscow, Russia}
\newcommand{\AFFILBC}{Institute for Theoretical and Experimental Physics,
117218~Moscow, Russia}
\newcommand{\AFFILBD}{Department of Physics, University of Cyprus,
1678~Nicosia, Cyprus}
\newcommand{\AFFILBE}{Institut de Physique Nucl\'{e}aire (UMR 8608), CNRS/IN2P3
- Universit\'{e} Paris Sud, F-91406~Orsay Cedex, France}
\newcommand{\AFFILBF}{Nuclear Physics Institute, Czech Academy of Sciences,
25068~Rez, Czech Republic}
\newcommand{\AFFILBG}{LabCAF. F. F\'{\i}sica, Univ. de Santiago de Compostela,
15706~Santiago de Compostela, Spain}
\newcommand{\AFFILCa}{ISEC Coimbra, Coimbra, Portugal}
\newcommand{\AFFILCb}{ExtreMe Matter Institute EMMI, 64291~Darmstadt, Germany}
\newcommand{\AFFILCc}{Technische Universit\"{a}t Dresden, 01062~Dresden,
Germany}
\newcommand{\AFFILCd}{Frederick University, 1036~Nicosia, Cyprus}
\newcommand{\AFFILCe}{Dipartimento di Fisica and INFN, Universit\`{a} di
Torino, 10125~Torino, Italy}
\newcommand{\AFFILCt}{Istituto Nazionale di Fisica Nucleare - Laboratori Nazionali del Sud, 95125~Catania, Italy}
\newcommand{\AFFILCr}{Moscow Engineering Physics Institute (State University), 115409 Moscow, Russia}
\newcommand{\AFFILCp}{Utrecht University, 3584 CC Utrecht, The Netherlands}
\newcommand{\AFFILGA}{NRC "Kurchatov Institute", PNPI, 188300, Gatchina, Russia}
\newcommand{\AFFILBN}{Helmholtz--Institut f\"ur Strahlen-- und Kernphysik, Universit\"at Bonn, Germany}

\author{J.~Adamczewski-Musch} \affiliation{\AFFILAD}
\author{O.~Arnold} \affiliation{\AFFILAI} \affiliation{\AFFILBJ}
\author{E.~T.~Atomssa} \affiliation{\AFFILBE}
\author{C.~Behnke} \affiliation{\AFFILAH}
\author{A.~Belounnas} \affiliation{\AFFILBE}
\author{A.~Belyaev} \affiliation{\AFFILAG}
\author{J.~C.~Berger-Chen} \affiliation{\AFFILAI} \affiliation{\AFFILBJ}
\author{J.~Biernat} \affiliation{\AFFILAC}
\author{A.~Blanco} \affiliation{\AFFILAB}
\author{C.~Blume} \affiliation{\AFFILAH}
% ...
\collaboration{HADES collaboration} \noaffiliation 
\author{A.~V.~Sarantsev} \affiliation{\AFFILGA} \affiliation{\AFFILBN}

\begin{abstract}
We report on the investigation of $\Delta$(1232) production and decay in proton-proton collisions at a kinetic energy of 1.25~GeV measured with HADES. Exclusive dilepton decay channels $ppe^{+}e^{-}$ and $ppe^{+}e^{-}\gamma$ have been studied and compared with the partial wave analysis of the hadronic $pp\pi^{0}$ channel. They allow to access both $\Delta^+ \to p\pi^0(e^+e^-\gamma)$ and $\Delta^+ \to  pe^+e^-$ Dalitz decay channels. The perfect reconstruction of the well known $\pi^0$ Dalitz decay serves as a proof  of the consistency of the analysis. The $\Delta$ Dalitz decay is identified for the first time and the sensitivity to N-$\Delta$ transition form factors is tested. The $\Delta$(1232) Dalitz decay branching ratio is also determined for the first time; our result is (4.19 $\pm$ 0.62 syst. $\pm$ 0.34 stat.) $\times $ 10$^{-5}$, albeit with some model dependence.
\end{abstract}


\maketitle

\section{Introduction}

One of the key issues in exploring the nature of the strong interactions over the years is the investigation of baryon resonances, i.e. short-lived excited states of nucleons. Its composite nature is probed in scattering experiments and is characterized not only by the complex pole position of the scattering amplitude but also by the couplings to the various channels and hence decay branching ratios. The electromagnetic structure of baryons is encoded in form factors and can be probed in two kinematical regimes defined by the sign of $q^2$ (four-momentum transfer squared) of the virtual photon: $q^2 <$  space-like, $q^2 >$ 0 time-like. In the space-like region, high precision experiments of electron and photon scattering delivered accurate data sets on $\gamma^{*}N \rightarrow N^{*}(\Delta)$ excitations for several resonances \cite{AznauryanBurkert2012,Aznauryan2013}. The time-like electromagnetic structure of baryonic transitions can also be studied in low-energy nucleon and pion induced collisions via $N^{*}(\Delta) \to N\gamma^{*} \to Ne^{+}e^{-}$ Dalitz decays. Due to the small positive four-momentum transfer squared ($q^{2} =  m^{2}_{\gamma^{*}}$), which is best suited to study the coupling to vector mesons, Dalitz decays give an insight into the "kinematically forbidden" time-like region, which is inaccessible in annihilation experiments. In the low-energy range, perturbative QCD cannot be applied and the understanding of baryon transitions is associated with the question about relevant degrees of freedom of these composite objects \cite{Lutz2016}. Although at higher four-momentum transfer the respective degrees of freedom might be considered effectively as constituent quarks, at lower four-momentum transfer, $q^2 < $ 1 GeV$^2$, besides a quark core also a meson cloud surrounding the quark core plays an important role \cite{Pasquini2006}. The coupling of virtual photons to hadrons is strongly affected in this regime by the light vector mesons and provides the foundation of Vector Meson Dominance model (VDM) \cite{Sakurai1960}.

\subsection{\bf{$\Delta$ properties and electromagnetic form factors}}
\label{delta_properties}
The $\Delta$(1232) resonance dominates pion production in $NN$ reactions for $\sqrt{s}~<$ 2.6 GeV/c$^2$. Despite its relatively large width (117 MeV) it is quite well separated from higher lying resonances. The dominating decay channel $\Delta \rightarrow N\pi$ has a branching ratio of 99.4\%, while the only measured electromagnetic decay $\Delta \rightarrow N\gamma$ has a branching ratio of 0.55-0.65\% \cite{PDG2016}. For the unmeasured $\Delta \to  N \gamma^{*}$ transition, a theoretical estimate on the level of 4$~\times~$10$^{-5}$ has been given \cite{Zetenyi2003}. The electromagnetic transition $N \to \Delta$ is predominantly magnetic dipole (M1) involving a spin and isospin flip of a single quark in the S-wave state. A small D-wave admixture of quadrupole (electric E2 and Coulomb C2) amplitudes describes small deformations of the resonance \cite{Aznauryan2009}. Electromagnetic decays can be parametrized by three helicity amplitudes $A_{1/2}(q^2)$, $A_{3/2}(q^2)$ and $S_{1/2}(q^2)$, defined in the $\Delta$ rest frame. The first two of them are related to the transverse photon polarization, the last one is related to a virtual longitudinal photon polarization. In the limit of a real photon ($q^{2}$ = 0), the amplitude $S_{1/2}$ vanishes. The best determinations of the helicity amplitudes $A_{1/2}$, $A_{3/2}$ for the real photon coupling were obtained in pion photoproduction experiments by the CLAS \cite{Workman2012}, MAMI/A2 \cite{Beck1997,Beck2000} and LEGS \cite{Blanpied1997,Blanpied2001} Collaborations. 

These helicity amplitudes are completely unknown for $q^2 > 0$. This region can be accessed via the Dalitz decay $\Delta \to  Ne^+e^-$. The differential decay width $d\Gamma$ can be expressed in terms of the resonance decay width $\Gamma^{\Delta \to N\gamma^*}$
\begin{equation}
\centering
\frac{d\Gamma^{\Delta \to Ne^+e^-}_{M_{\Delta}}}{dM_{ee}} = \frac{2\alpha}{3\pi M_{ee}}\Gamma^{\Delta \to N\gamma^{*}}_{M_{\Delta}}(M_{ee}),
\label{e_FACTOR}
\end{equation}hence it is also related to the radiative width $\Gamma^{\Delta \to N\gamma}_{M_{\Delta}}$. The calculation of the partial decay width $\Gamma_{e^{+}e^{-}N}$ requires the knowledge of the evolution of the electromagnetic transition form factors (eTFF) as a function of $q^2$, which are real in the space-like region, but get an imaginary part in the time-like region. They can be equivalently expressed in terms of the $\gamma^{*}N\Delta$ form factors: magnetic dipole ($G^{*}_{M}$), electric quadrupole ($G^{*}_{E}$) and Coulomb quadrupole ($G^{*}_{C}$), related to the discussed above helicity amplitudes, as introduced by Jones and Scadron \cite{JonesScadron1973}. The formula for the $\Gamma^{\Delta \to N\gamma^{*}}_{M_{\Delta}}$, derived by Krivoruchenko and F\"{a}ssler \cite{Krivoruchenko2002b}, has been applied in various model calculations \cite{IachelloWan2004,IachelloWan2005,Wan2007,RamalhoPena2012}. In the calculations of Z{\'e}t{\'e}nyi and Wolf \cite{Zetenyi2003} an equivalent set of form factors has been used, giving a consistent result. However, as pointed out by Krivoruchenko in \cite{Krivoruchenko2002b}, many former expressions for the Dalitz decays of baryonic resonances \cite{Wolf1990,Ernst1998} were inconsistent even in the real photon decay limit, i.e. $\Delta \to $ N$\gamma$. 

\subsection{\bf{eTFF of the $\Delta$ in model description}}
\label{eTTF_models}
In the past, a few models were proposed for the description of the $\Delta$ form factors. The "QED point-like" model of $\gamma^{*}N\Delta$ vertex \cite{Zetenyi2003,Dohrmann2010} incorporates the simplest constant form factors fixed from reactions with a real photon at $q^{2} = 0$. This assumption is based on the small four-momentum transfers involved in the $\Delta$ Dalitz decay and small values of $G_{E}$ and $G_{C}$ in the space-like region, as reported in \cite{Pascuta2007}. It results in $G_{M}$ = 3, $G_{E}$ = 0, $G_{C}$ = 0 and provides the correct radiative decay width $\Gamma^{\Delta \to N\gamma}$ = 0.66 MeV and BR($\Delta \to Ne^+e^-$) = 4.19~$\times$~10$^{-5}$ \cite{Zetenyi2003,Dohrmann2010} at the resonance pole, very close to $\alpha \times $BR($\Delta \rightarrow $ N$\gamma$). 

Another approach, the extended Vector Meson Dominance model (eVMD) by Krivoruchenko and Martemyanov \cite{Krivoruchenko2002b,Krivoruchenko2002a} includes the excited states of the vector mesons $\rho'$, $\rho''$,... etc. for the description of the eTTF, hence providing a more complete picture of the vector meson contribution to resonance decays. Parameters are constrained by the quark counting rules and photo- and electro-production amplitudes measured in the space-like region, as well as the decay amplitudes of nucleon resonances into a nucleon and a vector meson. 

The two component quark model by Iachello and Wan \cite{IachelloWan2004,IachelloWan2005,Wan2007} parametrizes the electromagnetic interaction with a direct and a vector meson coupling according to VDM. The dominant contribution (99.7\%) to the $G_{M}$ form factor is estimated by the VDM in terms of the dressed $\rho$-meson propagator, being dominant for the range of $q^{2}$ involved in the Dalitz decays. The Iachello-Wan model was the first model of the eTFF which was analytically extended to the time-like region. It was very successful in describing the existing data of nucleon form factors in the space-like region. However, it used the pole position of the $\rho$-meson at a significantly lower value than expected \cite{Dohrmann2010}. In addition, a comparison with the dilepton data collected by HADES in proton-proton collisions at a higher kinetic beam energy (3.5 GeV) \cite{Przygoda2014} unravels that the $\Delta$(1232) contribution with the Iachello-Wan form factor parametrisation can describe inclusive $e^+e^-$ spectra well \cite{Weil2012} but leaves no space for the expected contributions of the higher resonances. 

The most recent covariant constituent quark model by Ramalho-Pe\~{n}a \cite{Ramalho2016} provides the description of the dominant $G^{*}_{M}$ by means of two contributions: the quark core and the pion cloud dressing. The quark core component \cite{Gross1969,GrossRamalho2008} describes the resonance as a quark-diquark structure as an $S$-wave state (the electric $G^{*}_{E}$ and Coulomb $G^{*}_{C}$ quadrupole form factors originate from the small $\leq 1\%$ admixture of a $D$ state \cite{Ramalho2008b}). The valence quark component is determined from the lattice QCD and in agreement with the data in the space-like region (the EBAC analysis of pion photoproduction) \cite{Ramalho2009a,Ramalho2009b}. The comparison with data allows also for the extraction of the meson cloud component in the space-like region \cite{RamalhoPena2012,Ramalho2008}. However, the model description of the time-like region requires an analytical extension. The contribution of the pion cloud to $G^{*}_{M}$ is parametrized with two terms: a photon directly coupling to a pion or to intermediate baryon states. The parametrization of the pion eTFF, used in the coupling, is in agreement with the available high-precision data \cite{Ramalho2016}. One should note that, in contrast to \cite{IachelloWan2004}, this parametrization takes properly into account the $\rho$-meson pole and width. In consequence, as shown in \cite{Ramalho2016}, the inclusive HADES data on $e^+e^-$ production in $NN$ collisions at 2.2~GeV \cite{Hades2012} and 3.5~GeV \cite{Hades2012b} are well described, including contributions from higher mass resonances. In the $q^2 < $ 0.3 GeV$^2$ region, relevant for this study, the model predicts a dominant contribution of the pion cloud, increasing as a function of the mass, and an almost constant contribution from the quark core. 

\subsection{\bf{Nucleon-nucleon bremsstrahlung}}
\label{nn_bremss}
Another source of virtual photons and hence $e^+e^-$ pairs is the nucleon-nucleon bremsstrahlung produced in the strong interaction field of two nucleons without intermediate resonance excitation. A description of this process combines the $NN\gamma^{*}$ electromagnetic vertex and the $NN$ interaction. One possibility to describe this process is the Soft Photon Approximation (SPA) \cite{Wolf1990,Gale1989} which, however, is limited to low-energy photons due to the on-shell treatment of $NN\gamma^{*}$ vertex and lacks of emission off internal exchange boson.

The final state of $ppe^+e^-$ or $npe^+e^-$ may result in our energy range from both $\Delta$ Dalitz decay and $NN$ bremsstrahlung, and both processes can interfere. The coherent sum of the contributing amplitudes has been evaluated within the One-Boson Exchange (OBE) models, e.g. by Kaptari and K\"{a}mpfer \cite{Kaptari2009} and Shyam and Mosel \cite{Shyam2010}. These models provide the total $e^+e^-$ contribution based on a coherent treatment of many amplitudes including contributions of the $\Delta$ resonance and the nucleon-nucleon bremsstrahlung. Despite the similar approaches the models give different predictions for the energy dependence of differential cross sections. One should note that according to both OBE models, interference effects between the $\Delta$ and $NN$ bremsstrahlung are small and play a role only at the higher $e^+e^-$ invariant mass ($M_{e^+e^-} >$ 0.4 GeV/c$^2$). This seems to justify the statement that in proton-proton collisions the $NN$ bremsstrahlung contribution can be treated separately and added incoherently to the $\Delta$ contribution.

\subsection{Exploring resonance production by HADES}
\label{sec_resonance}
The High-Acceptance Di-Electron Spectrometer (HADES) is a versatile magnetic spectrometer \cite{Hades2009} installed at SIS18, GSI Darmstadt. Thanks to its high acceptance, powerful particle (p/K/$\pi$/e) identification and very good mass resolution ($2-3\%$ for dielectrons in the light-vector meson mass range) it allows the study of both hadron and rare dielectron production in $N+N$ \cite{Hades2010,Hades2012,Hades2012b,Hades2014}, $p+A$ \cite{Hades2012pNb}, $A+A$ \cite{Hades2007CC,Hades2008CC,Hades2011}, $\pi+p$ \cite{Hades2016a,Hades2016b}, $\pi+A$ collisions in the beam energy range of a few (A)GeV. Nucleon-nucleon reactions play a special role in this context, providing an important reference for $p+A$ and $A+A$ collisions.

The proton beam energy of 1.25~GeV was selected below the $\eta$ meson production threshold in order to favour $\Delta$(1232) production. It was extensively studied via exclusive channels with one pion in the final states $np\pi^{+}$ and $pp\pi^{0}$ by HADES. The first attempt based on the resonance model of Teis \textit{et al.} \cite{Teis1997} unravelled difficulties in the description of both the yield and angular projections \cite{Ramstein2012}. Extended studies based on various observables in the framework of a partial wave analysis (PWA) of the Bonn-Gatchina group \cite{Ermakov2011} provided much better description and confirmed the dominant contribution of the $\Delta$(1232), yet with a sizable impact of $N$(1440) and non-resonant partial waves \cite{Przygoda2015}. The results pave the way to studies of $\Delta^{+}$ resonance measured in the same experiment in the dielectron channels ($ppe^+e^-\gamma$ and $ppe^+e^-$) with a focus on the resonance Dalitz decay which has never been measured before.
\\
\\
Our paper is organized as follows. Section \ref{thehadesexp} introduces the experimental  conditions under which $ppe^+e^-\gamma$ and $ppe^+e^-$ channels were selected and the normalization procedure. The $\pi^{0}$ Dalitz decay is identified (Sec.~\ref{fourprong}) in the $ppe^+e^-\gamma$ final state and various differential distributions (acceptance corrected) are compared to the PWA description. The total production cross section for the $\pi^{0}$ production is deduced and compared to the one obtained from hadronic channel. Section \ref{threeprong} presents the identification of the final state of $ppe^+e^-$. Invariant masses $e^+e^-$ and $pe^+e^-$ are discussed within the HADES acceptance as well as the acceptance corrected angular distributions. The data are confronted with models describing the eTFF of $\Delta$: a point-like $\gamma^{*}NR$ coupling and a covariant constituent quark model in Section \ref{pepem_mc}. The non-resonant virtual photon emission is also discussed and estimated. Finally, in Section \ref{delta_br}, the $\Delta$ Dalitz decay branching ratio is determined. We summarize in Section \ref{endsummary}.

\section{Proton-proton experiment and analysis}	
\label{thehadesexp}
A proton beam of 10$^7$ particles/s was incident on a liquid hydrogen target with a length of 5 cm (total thickness of $\rho d$~=~0.35~g/cm$^{2}$). The data readout was started upon a first level trigger (LVL1) decision with the charged particle multiplicity $\geq$ 3 with all events written to the tape. The LVL1 condition was followed by a second level trigger (LVL2) requesting at least one lepton track candidate to record events of the type $e^{+}e^{-}X$. The LVL2 trigger efficiency amounts to 0.84 and it has been deduced in Monte Carlo simulations to be independent of the $e^{+}e^{-}$ pair mass. 

\subsection{\bf{Particle identification and time reconstruction}}
The following event hypotheses were studied in this paper: 4-prong ($ppee$) and 3-prong ($pee$) analyses. They require the identification of protons, electrons and positrons. The first step of the analysis (lepton and hadron identification, track reconstruction) is described in detail in \cite{Hades2009}. Each track was required not to have any partially or fully reconstructed neighbouring track within an angle of 5$^{\circ}$ in order to reduce fake or double (ghost) particle reconstruction. In the absence of the START detector \cite{STARTdetector2010} only relative time-of-flight of particles, measured in the two detectors (TOF, TOFino) in a given event was available. In conjunction with the reconstructed momentum it was possible to build all possible particle combinations (hypotheses) out of the pool of hadronic and leptonic tracks, with positive or negative charge. A graphical particle identification (PID) two-dimensional cut (momentum vs effective mass squared) was derived from the experimental data compared to Monte Carlo simulations. It served to select the given event hypothesis ($ppe^+e^-$ or $pe^+e^-$) with the lowest $\chi^{2}$ (taking into account time resolution $\sigma_{TOF}\sim $ 150 ps and $\sigma_{TOFino}\sim $ 450 ps). To calculate the time-of-flight for all particles in the event one particle has to be defined as a reference particle, the optimum selection being an electron. The latter is identified with a high purity by the RICH detector in HADES \cite{Hades2009}. The time reconstruction procedure introduces a systematic error lower than 2\% in the $e^+e^-$ signal yield.

\subsection{\bf{Dielectron signal selection}}
\label{dilepton_signal}
For reactions with final $e^{+}e^{-}$ pairs, the combinatorial background (CB) was obtained using the geometric mean 
\begin{equation}
\frac{dN_{CB}}{dM_{e^{+}e^{-}}}=2\sqrt{\Big(\frac{dN}{dM}\Big)_{e^{+}e^{+}}\Big(\frac{dN}{dM}\Big)_{e^{-}e^{-}}},
\label{e_CB}
\end{equation}
where $e^{+}e^{+}$ and $e^{-}e^{-}$ stand for the same-event like-sign pairs. This allowed to account for the correlated background from the $\gamma$ conversion (mostly $\gamma$ from $\pi^{0}$ decays) as well as uncorrelated background from multi-pion decays. All distributions (invariant masses, angular projections) built upon the $e^{+}e^{-}$ pairs will be presented with the CB subtracted. To suppress  the conversion contribution, an opening angle larger than 9$^{\circ}$ between lepton tracks was required for both unlike-sign and like-sign pairs. 

\subsection{\bf{Analysis strategy and normalization}}
The following $p+p$ reaction dilepton channels are discussed:
\begin{itemize}
 \item $\pi^{0} \rightarrow $ e$^{+}$e$^{-}\gamma$ (BR = 1.194$\times$10$^{-2}$): The identification of the $\pi^{0}$ in a four-prong channel ($ppe^+e^-\gamma$) allows for the comparison of various differential distributions with the ones extracted from the hadronic channel $pp\pi^0$ \cite{Przygoda2015}. 
 \item $\Delta^{+} \rightarrow $ pe$^{+}$e$^{-}$ (theoretical estimate of BR = 4.2$\times$10$^{-5}$ \cite{Zetenyi2003}) for the invariant mass $M_{e^{+}e^{-}}>$~0.14~GeV/c$^{2}$. The baryonic resonance is identified based on selected characteristic distributions, the $\Delta$ angular production and the decay and $\Delta$ invariant mass distributions.
\end{itemize}
All presented spectra (if not stated otherwise) were normalized to the p+p elastic scattering yield measured in the same experimental run. The reference p+p elastic cross section for the proton in the polar angle range between 46$^{\circ}$-134$^{\circ}$ in the c.m.s. amounts to 3.99~$\pm$~0.19 mb (EDDA Collaboration \cite{EDDA2004}). The normalization error is estimated to be 8\%, where 5\% is derived from the error of the reference differential cross section and 6\% is the systematic error of the reconstruction of events with elastic scattering in HADES (see \cite{Ramstein2012} for details).

\begin{figure*}
  \centering
    \includegraphics[width=0.49\textwidth]{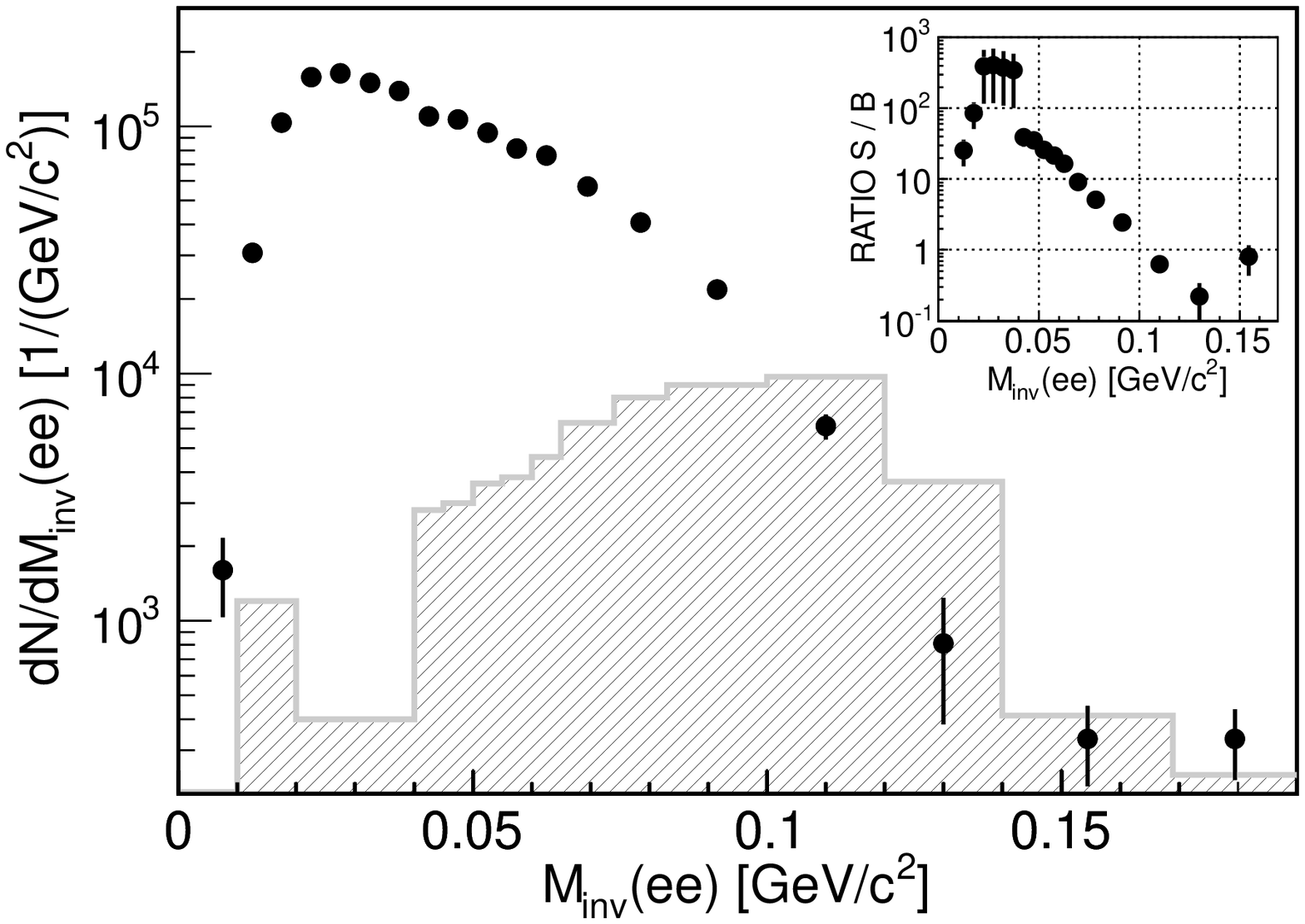}
    \includegraphics[width=0.49\textwidth]{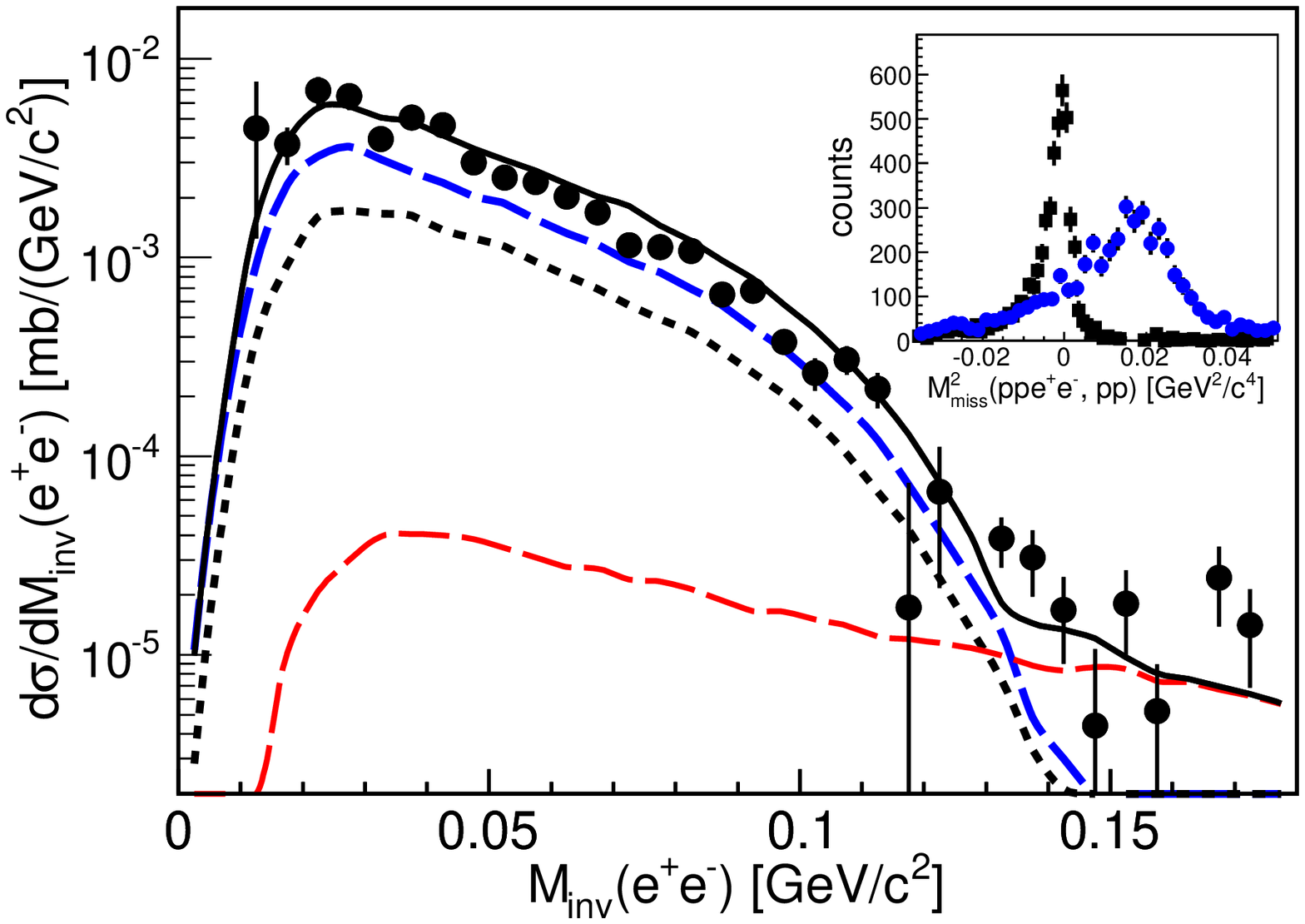}  
  %\vspace*{-0.2cm}
  \caption {(Color online) $ppe^{+}e^{-}\gamma$ final state. Left: $e^{+}e^{-}$ uncorrected invariant mass distribution of signal pairs (number of counts is per GeV/c$^2$ to account for the variable bin width) with the cuts selecting $\pi^{0}$. A gray hatched area represents the combinatorial background (CB). Inset: signal to background ratio. Right: $e^{+}e^{-}$ invariant mass within the HADES acceptance. Experimental data (black dots) are corrected for the detection and reconstruction efficiencies. Normalization error is not indicated. The curves display Monte Carlo simulations. Black solid curve represents the total $\pi^{0}$ Dalitz decay as obtained in the partial wave analysis \cite{Przygoda2015}. In addition, decomposition to resonances ($\Delta$ - blue dashed curve, $N(1440)$ - black short dashed curve) decaying to $p\pi^{0}$; the components are added coherently. Red dashed curve represents $\Delta$ Dalitz contribution in a description with a point-like $\gamma^{*}NR$ coupling \cite{Zetenyi2003,Dohrmann2010}. Inset: missing mass squared of $ppe^+e^-$ - black points and missing mass squared of two protons - blue points. 
    }
  %\vspace*{-0.2cm}
  \label{ppepem_rawdata}
\end{figure*}

The data are compared to various theoretical models. The model contributions are always obtained based on the full GEANT simulation (with implemented spectrometer geometry and materials of the subdetectors) and the Monte Carlo simulations of the detector response to charged particles, followed by the channel selection (hit identification, track reconstruction) likewise in the analysis of the experimental data.

\section{\bf{$ppe^+e^-\gamma$} channel and \bf{$\pi^{0}$} identification}
\label{fourprong}
The production of $\pi^{0}$ has been studied in the analysis of channels with one pion in the final state \cite{Przygoda2015}. Out of the two dominant decay channels ($\pi^{0} \rightarrow  \gamma\gamma$ and $\pi^{0} \rightarrow \gamma e^{+}e^{-}$) the latter one, the $\pi^{0}$ Dalitz decay, can be completely reconstructed with the missing mass technique by the identification of four particles p, p, e$^{+}$, e$^{-}$ in the $ppe^+e^-\gamma$ final state. 

Figure \ref{ppepem_rawdata} (left panel) shows the $e^+e^-$ invariant mass spectrum as the number of signal pairs ($e^+e^-$ pairs after combinatorial background subtraction) per GeV/c$^2$, to account for the variable bin size used. The combinatorial background is depicted as a gray hatched area. A strong increase in the CB near the $\pi^{0}$ mass signals the correlated source of dielectrons produced in the conversion of two real photons in the same event, following the $\pi^{0} \rightarrow \gamma\gamma$ decay. If both $e^+$ and $e^-$ produced  by the same photon are registered, the conversion is effectively suppressed by the $e^+e^-$ opening angle cut (see Sec.~\ref{dilepton_signal}). If only one track from each photon is reconstructed, it contributes to the combinatorial background. The signal-to-background ratio is very high, reaching the value of 400 (see the inset in Fig.~\ref{ppepem_rawdata}, left panel) and dropping down below 1 near $M_{e^+e^-} \sim$ 0.14 GeV/c$^2$. To provide a clean signal, a two-dimensional cut on the missing mass of two protons squared (where the missing particle is $\pi^0$) and the missing mass of four particles: p, p, e$^+$, e$^-$ squared (where the missing particle is $\gamma$) is applied with a window selecting 95\% of all events. Figure~\ref{ppepem_rawdata}, in right panel inset, shows the projected distributions of the missing masses squared. It has been checked both by the experimental data and the Monte Carlo simulation that the variation of the selection window width introduces a systematic error lower than 10\%. The number of reconstructed $e^+e^-$ pairs amounts to 7500. 

\begin{figure*}
       \centering
       \includegraphics[width=0.24\textwidth]{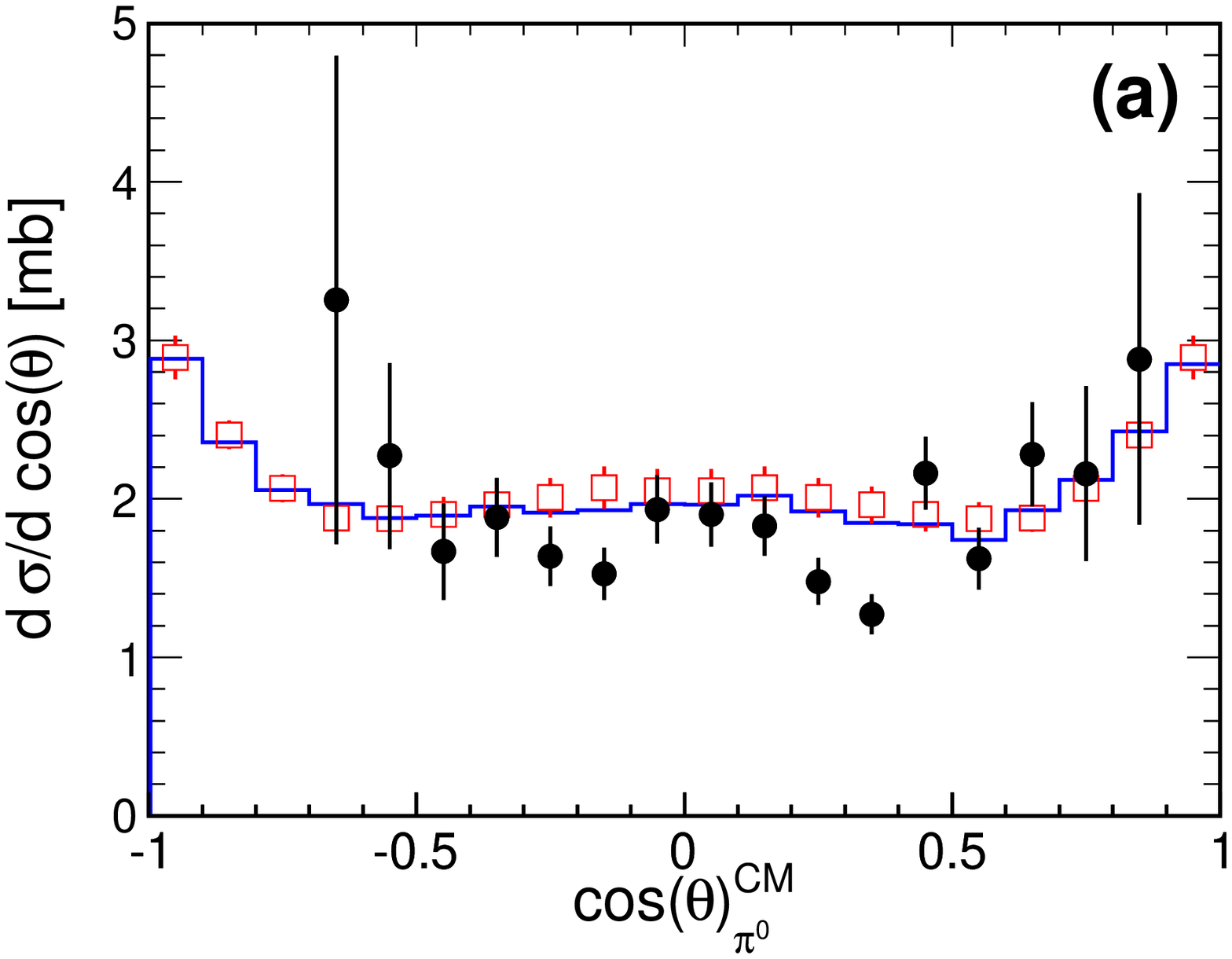}
       \includegraphics[width=0.24\textwidth]{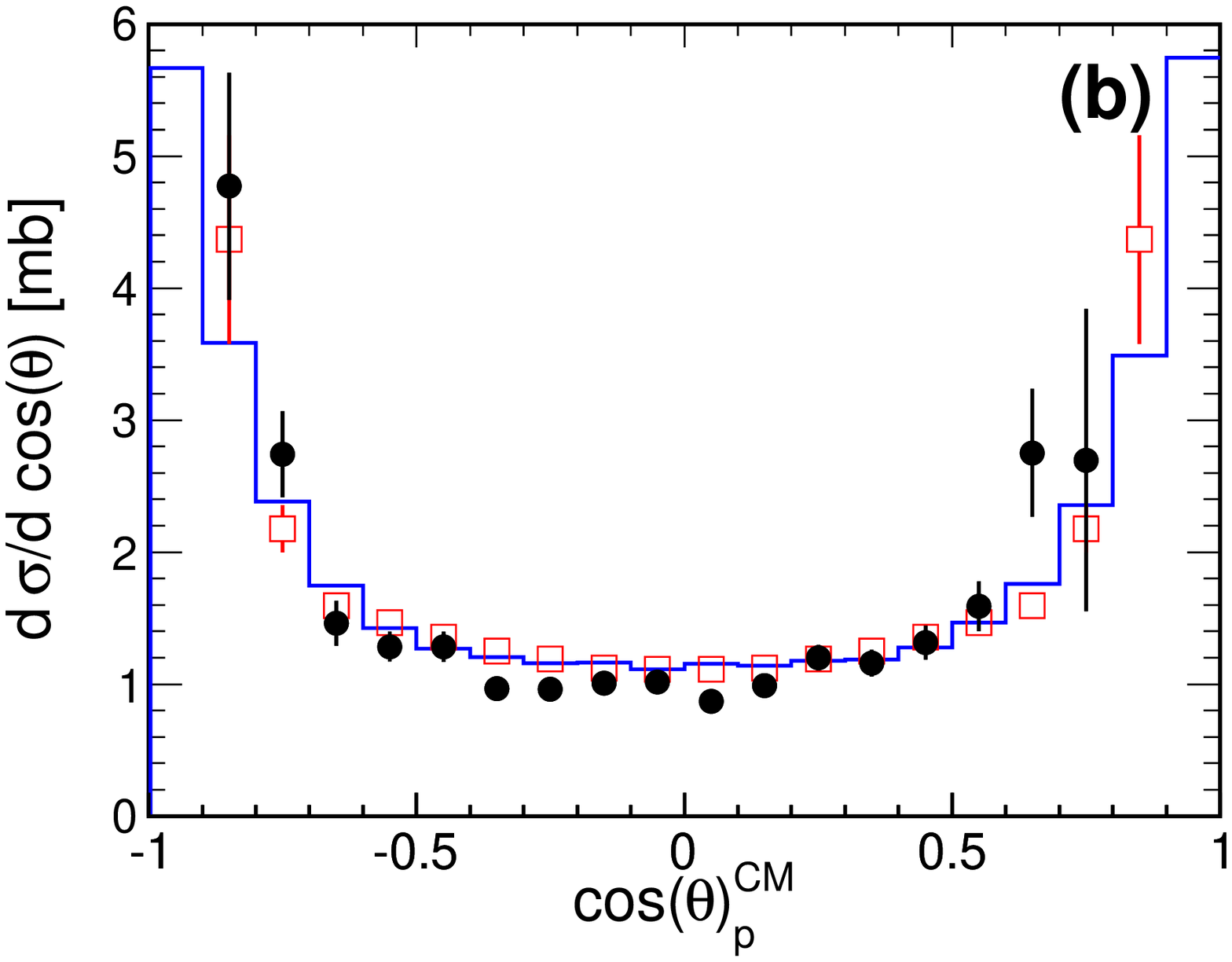}
       \includegraphics[width=0.24\textwidth]{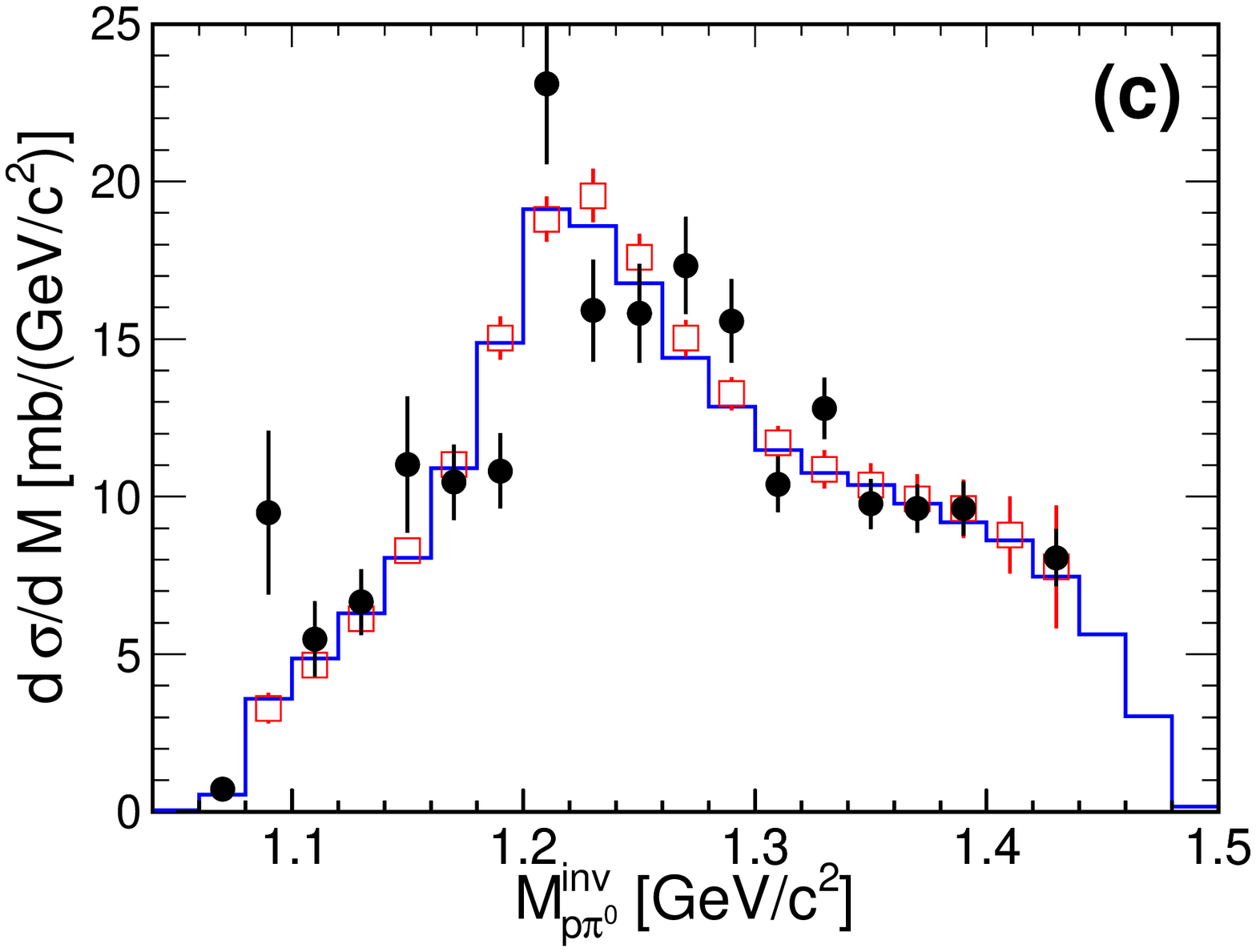}
       \includegraphics[width=0.24\textwidth]{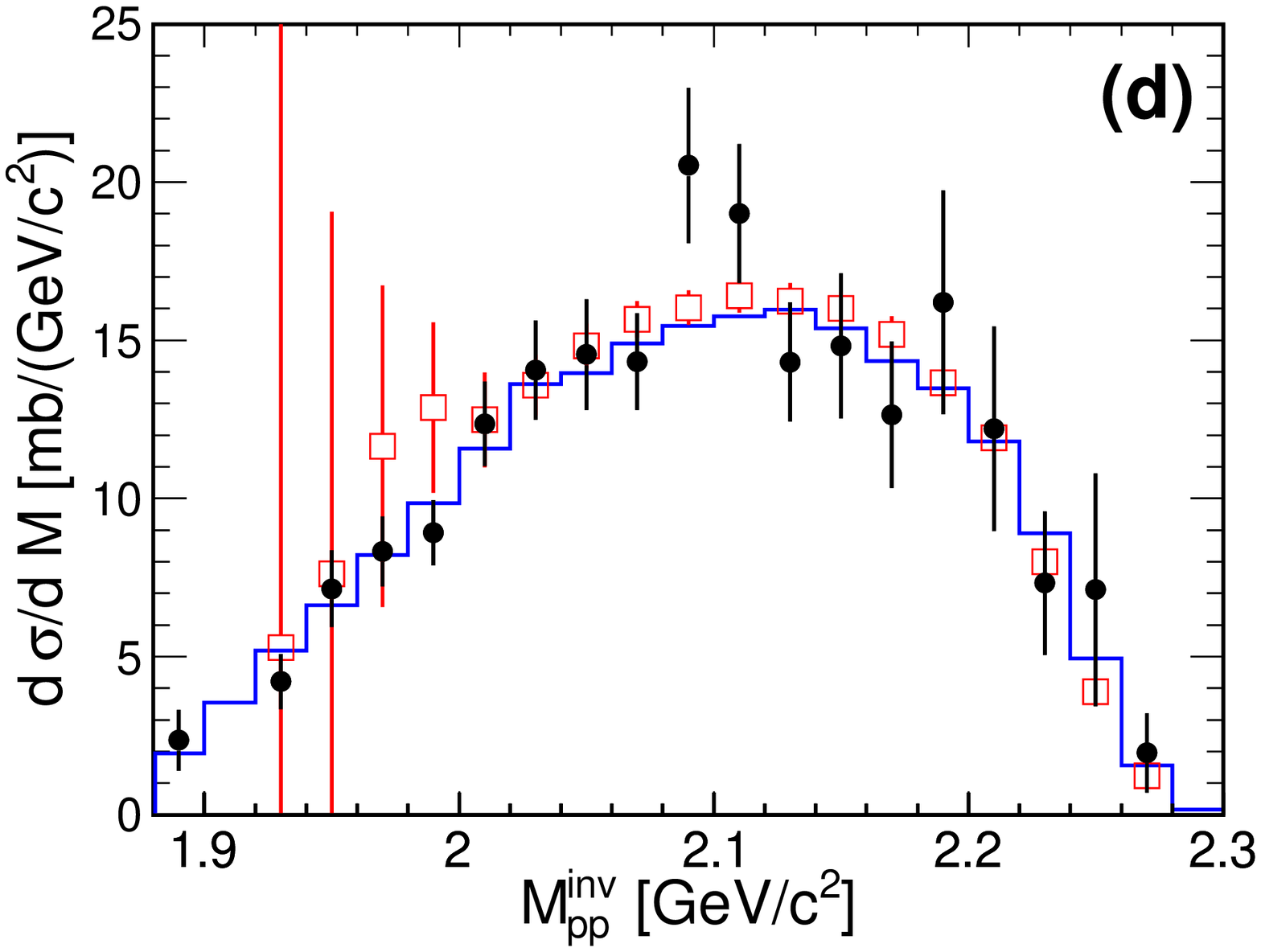}\\
       
       \includegraphics[width=0.24\textwidth]{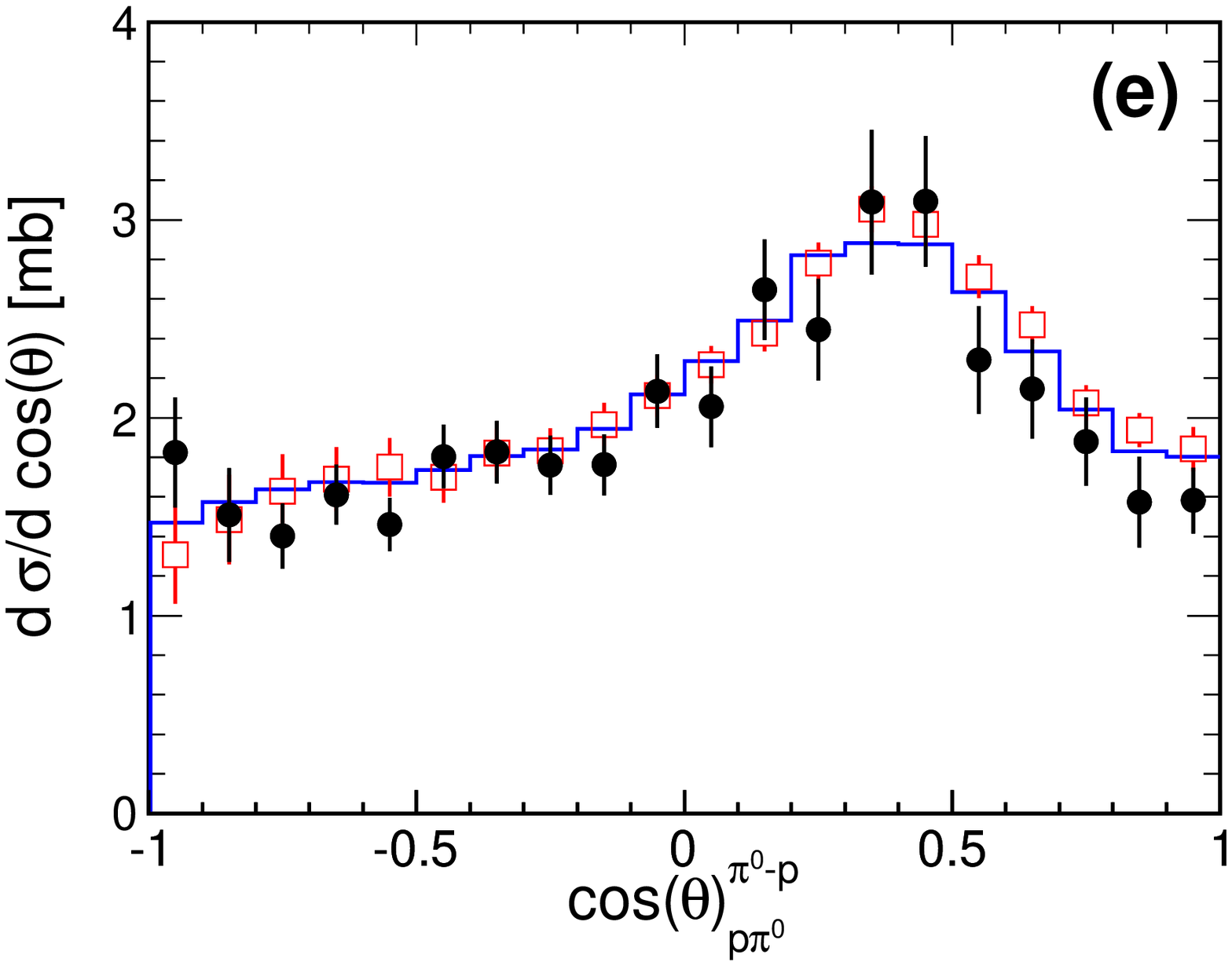}
       \includegraphics[width=0.24\textwidth]{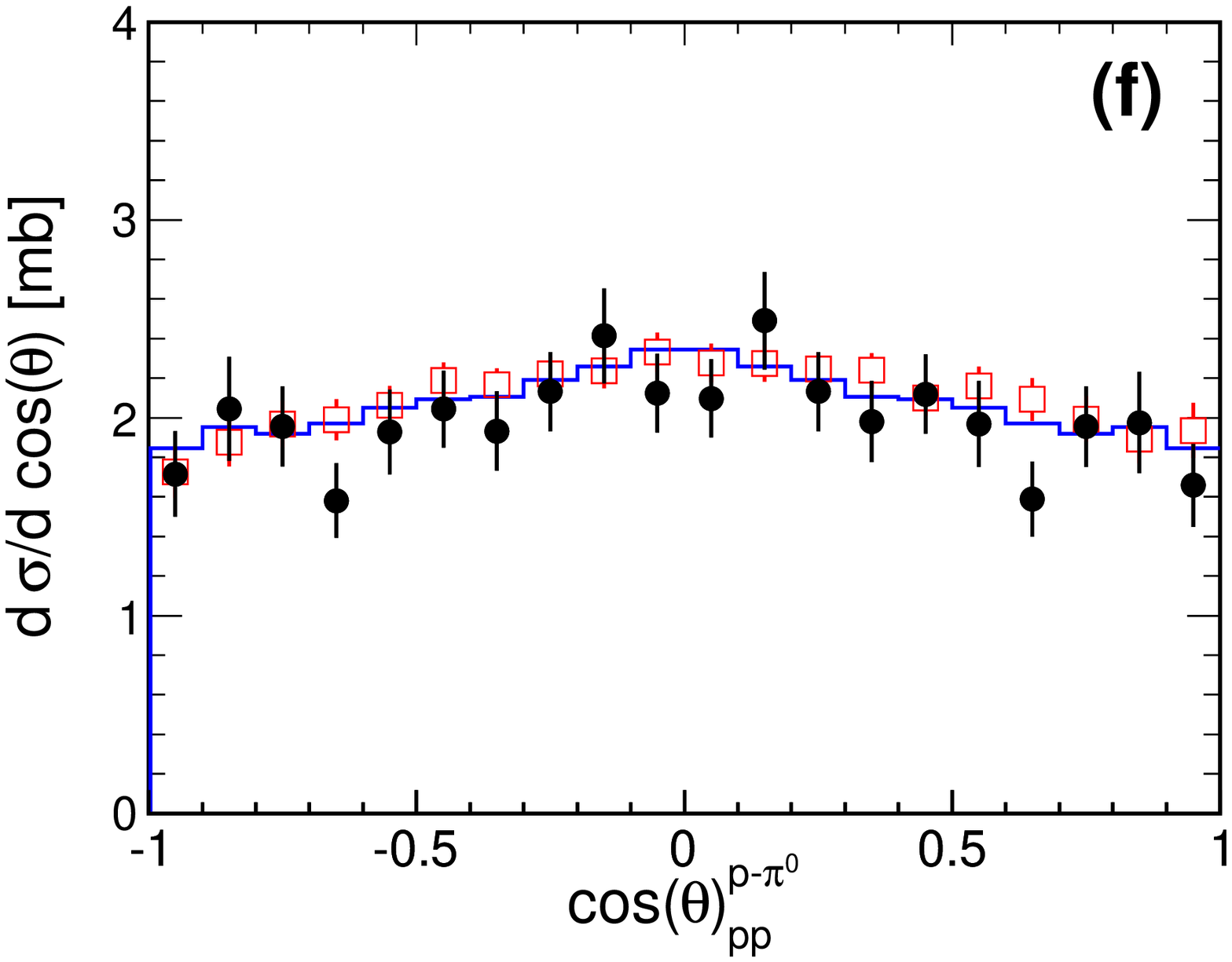}                    
       \includegraphics[width=0.24\textwidth]{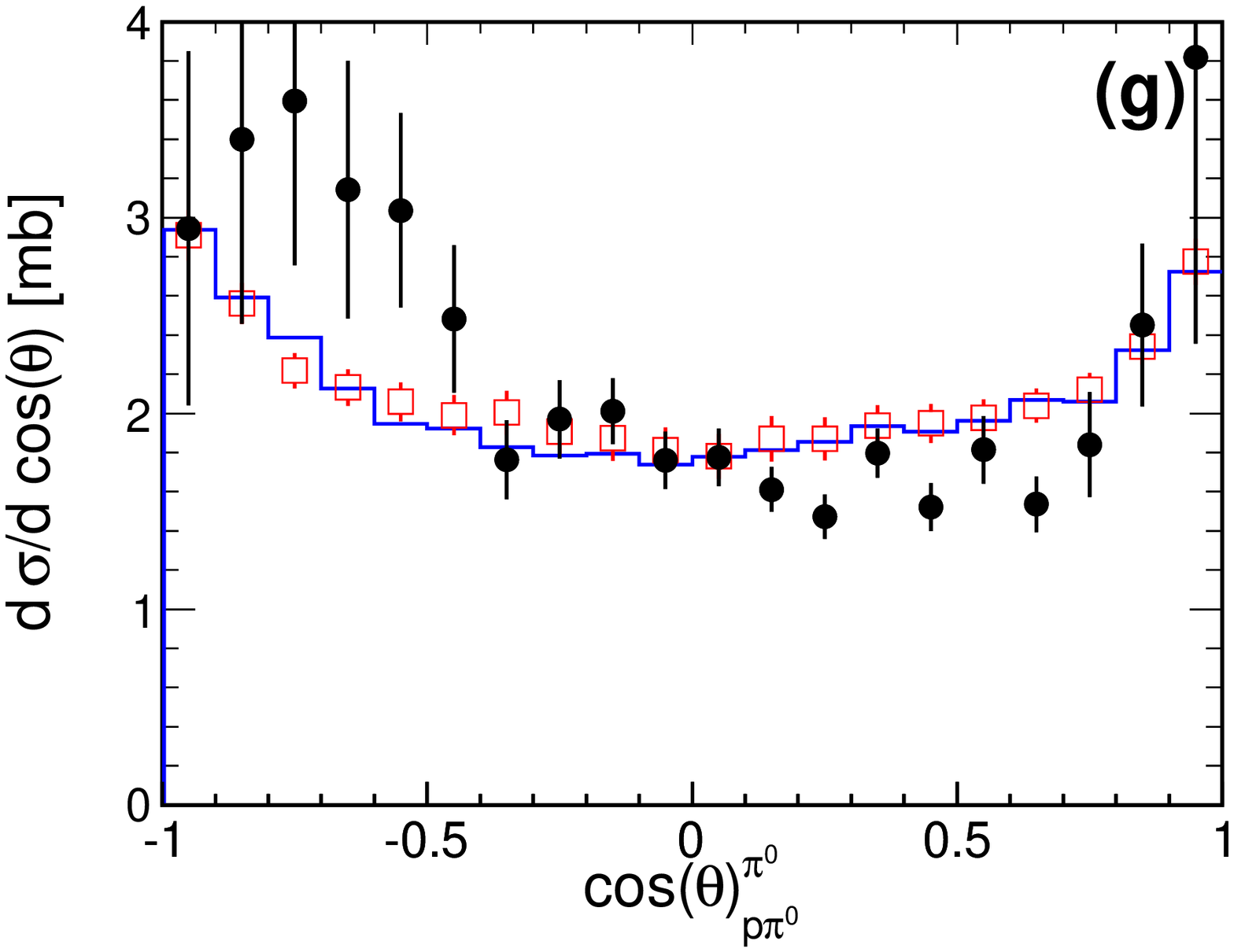}
       \includegraphics[width=0.24\textwidth]{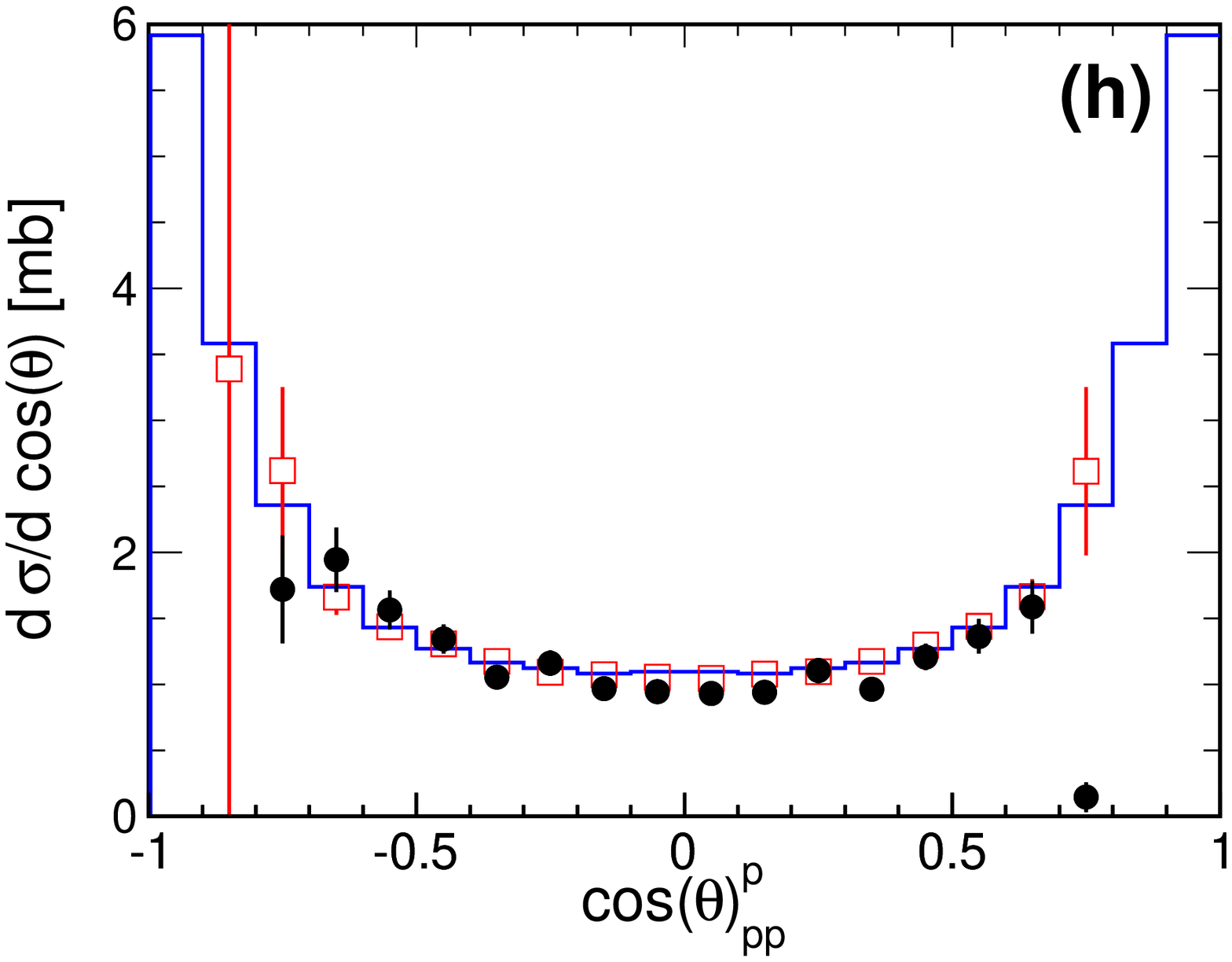}\\
       
       \caption{(Color online) Various projections of the $pp\pi^{0}$ and $pp\pi^{0} \to ppe^+e^-\gamma$ channels. Top row: angular distributions of (a) $\pi^{0}$ and (b) $p$ in c.m.s. reference frame; invariant mass distributions of (c) $p\pi^{0}$ and (d) $pp$. Bottom row: angular distributions in the helicity frame of (e) $\pi^{0}$ in $p\pi^{0}$ reference frame and (f) $p$ in $pp$ reference frame; angular distribution of (g) $\pi^{0}$ in $p\pi^{0}$ GJ reference frame and (h) $p$ in $pp$ GJ reference frame. Dielectron data points after acceptance and BR($\pi^{0} \rightarrow \gamma e^{+}e^{-}$) corrections (black dots) are compared with the data from hadron channel (red open squares). Histograms: total PWA solution (solid blue) obtained for the hadronic channel (see text for details).}
\label{ppepem_hadronic}
\end{figure*}

Figure \ref{ppepem_rawdata} (right panel) presents the invariant mass spectrum of $e^+e^-$ within the HADES acceptance. It has been corrected for the detection and reconstruction inefficiencies. The correction is done with the help of a one-dimensional correction histogram, deduced from the Monte Carlo simulations. The correction factor for the masses below $\pi^{0}$ mass amounts to about 20.

\subsection{\bf{Monte Carlo simulation and results}}
\label{ppepem_mc}
The partial wave analysis of the final state $pp\pi^{0}$ \cite{Przygoda2015} provided a very good description of this hadronic channel both in terms of the total cross section 4.2 $\pm$ 0.15 mb and the various differential distributions. This analysis provided $pp\pi^{0}$ events distributed according to the PWA solution. The $\pi^{0} \rightarrow e^+e^-\gamma$ decay was further implemented in order to generate the full $ppe^+e^-\gamma$ final state and to compare the experimental distributions with the PWA-driven simulated events. Figure \ref{ppepem_rawdata} (right panel) shows such a comparison for $M_{inv}(e^+e^-$) within the HADES acceptance. The systematic error of experimental data is 12\%. It includes the particle identification, the time reconstruction, the CB rejection and the missing mass selection. The statistical error is negligible in the $\pi^{0}$ region. The normalization error, given above, is not shown. The Monte Carlo simulation is shown in comparison. The black curve depicts the contribution from all $\pi^{0}$ Dalitz decay events, describing the data very satisfactorily. In addition, the decomposition to the intermediate resonance states is shown: blue dashed curve for $\Delta$(1232) and black short dashed curve for $N(1440)$, are given by the PWA solutions. There are also non-negligible non-resonant contributions (not shown in the picture). The amplitudes need to be added coherently in order to obtain the total $\pi^{0}$ contribution (black curve). $\Delta$ Dalitz decay is shown as a red dashed curve (for details on this contribution, see Sec.~\ref{pepem_mc}). 

For better verification of the dielectron channel in the $\pi^{0}$ Dalitz decay, various distributions of experimental data were compared with the PWA solution and hadron data as in \cite{Przygoda2015}. In Fig.~\ref{ppepem_hadronic}, we show single particle angular distributions in the center of mass (c.m.s.), helicity and Gottfried-Jackson (GJ) frames and two-particle invariant mass spectra. The data were corrected for the reconstruction efficiencies and the detector acceptance, each distribution with the respective one-dimensional correction function. The correction function is constructed, for a given distribution, as ratio of the model yield in the full solid angle (as provided by the PWA solution) and the yield within the HADES acceptance, including all detection and reconstruction efficiencies obtained using the full analysis chain. The correction factor in the $\pi^{0}$ Dalitz decay channel varies in the range 30-50. A direct comparison with the distributions for the hadronic channel requires a correction of dilepton data by the inverse of the BR = 1.194$~\times~10^{-2}$. All projections in Fig. \ref{ppepem_hadronic} demonstrate that the $\pi^{0}$ Dalitz decay reconstruction is well under control and both data in the dielectron and hadronic channels are well described by the PWA solution.

\begin{figure}[!htb]
  \centering
    \includegraphics[width=0.49\textwidth]{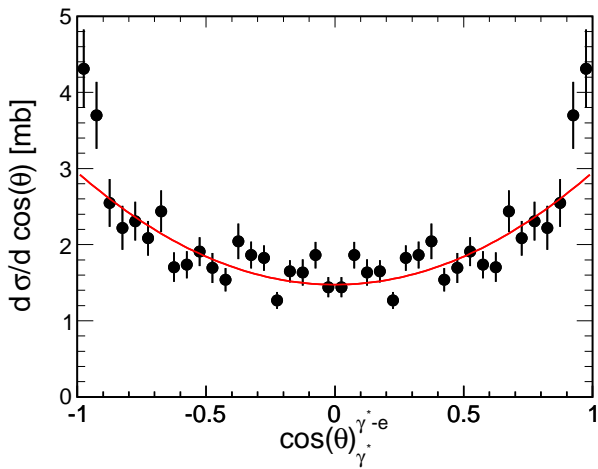}
  %\vspace*{-0.2cm}
  \caption {(Color online) $ppe^{+}e^{-}\gamma$ final state. The angle between $e^{+}$ or $e^{-}$ and $\gamma^{*}$ (upper index) in the $\gamma^{*}$ rest frame (lower index) for $\pi^{0}$ Dalitz decay (acceptance corrected). Experiment - black data points with statistical errors only. Fitted formula (red curve) $\propto 1+B\cos^{2}\theta$, with $B$~=~1.00~$\pm$~0.11, in agreement with the expected value (see \cite{Bratkovskaya1995}). The distribution is symmetrized by plotting both $e^{+}$ and $e^{-}$ contributions.
    }
  %\vspace*{-0.2cm}
  \label{ppepem_helicity}
\end{figure}

Yet another observable sensitive to the structure of the electromagnetic transition is defined as the angle between a lepton ($e^+$ or $e^-$) and the virtual photon $\gamma^{*}$ in the rest frame of $\gamma^{*}$, first boosted (leptons and $\gamma^{*}$) to the rest frame of the decaying resonance. This angular distribution has the form $1+B\cos^2\theta$ \cite{Bratkovskaya1995}. In the simplest case of scalar mesons ($\pi^{0}$, $\eta$), the anisotropy coefficient is 1, since the helicity conservation in the $\gamma^{*} \gamma$ decay allows, for the pseudoscalar-vector-vector transitions, only for transverse virtual photons. Figure~\ref{ppepem_helicity} presents the acceptance corrected $e^+$ or $e^-$ angle in the $\gamma^{*}$ reference frame in the reconstructed $\pi^{0}$ Dalitz decay channel. The distribution is symmetrized by plotting both $e^+$ and $e^-$ contributions. The fit (red curve) returns the parameter $B$ = 1.00 $\pm$ 0.11. In addition, the data are also corrected for the BR($\pi^{0} \rightarrow \gamma e^+e^-$) and the integral over the angular distribution results in the total cross section for the $\pi^{0}$ production, $\sigma$($pp~\rightarrow pp\pi^{0}$) = 4.18 mb. The statistics error is negligible (less than 2\%), the systematic and normalization errors are 12\% and 8\%, respectively, as discussed above. Both the anisotropy and the deduced cross section are in agreement with the predictions for the neutral pion Dalitz decay and the description of the $\pi^{0}$ production in the PWA framework ($\sigma^{PWA}_{\pi^0} = 4.2 \pm 0.15 $ mb). These results prove the perfect consistency of the analyses of the leptonic and hadronic channels for the $\pi^0$ reconstruction. On the one hand, it demonstrates the high quality of the reconstruction of electromagnetic channels with HADES which will be further exploited for the reconstruction of the $ppe^+e^-$ channel. On the other hand, it confirms the validity of the PWA analysis, providing the $\Delta^+$ contribution, which is essential for the BR($\Delta \to pe^+e^-$) estimate. 

According to the PWA description \cite{Przygoda2015} the contribution of the $\Delta$ resonance to the channel with one neutral pion in the final state is 70\%. The remaining part results from $N(1440)$ decay and non-resonant $^{3}P_{2}$ partial wave, destructively interfering with the Roper resonance. Since no notable influence of interferences with non-resonant partial waves was observed for the $\Delta$(1232) contribution, the estimate from the PWA can be safely taken as the $\Delta$ production cross section input for the simulation of the $\Delta$ Dalitz decay. In addition, the contribution of nucleon-nucleon bremsstrahlung is expected to be small, as it will be discussed in the next section.

\section{\bf{$ppe^+e^-$} channel and $\Delta$ identification}
\label{threeprong}
The identification of three particles ($p$, $e^{+}$, $e^{-}$) in the $ppe^+e^-$ final state allows for the kinematically complete (exclusive) reconstruction of the $\Delta$ Dalitz decay channel under two conditions: a) selection of the missing mass of $pe^{+}e^{-}$ ($M_{miss}^{pe^{+}e^{-}}$), close to the proton mass as a signature of the exclusive $pp~\rightarrow ppe^{+}e^{-}$ reaction; b) invariant mass $M_{e^{+}e^{-}} > M_{\pi^{0}}$ for a rejection of the $\pi^{0}$ Dalitz decay. Although the exit channel is in this case $\gamma$e$^+$e$^-$, it is only partially suppressed by cut a) due to the finite missing mass resolution and the cut b) is needed for the channel separation. In about 20\% of all events both protons are measured. Since there is no clear identification of the proton produced by the $\Delta$ decay, all projections using proton variables in their construction are added with a weight of 0.5 for both protons in the final state, i.e. a) both protons ($p_{1}e^+e^-$) and ($p_{2}e^+e^-$) if $p_{1}$ and $p_{2}$ are measured; b) measured proton ($pe^+e^-$) and missing proton ($p_{miss}e^+e^-$) if only $p$ is measured.

\begin{figure*}
  \centering
    \includegraphics[width=0.49\textwidth]{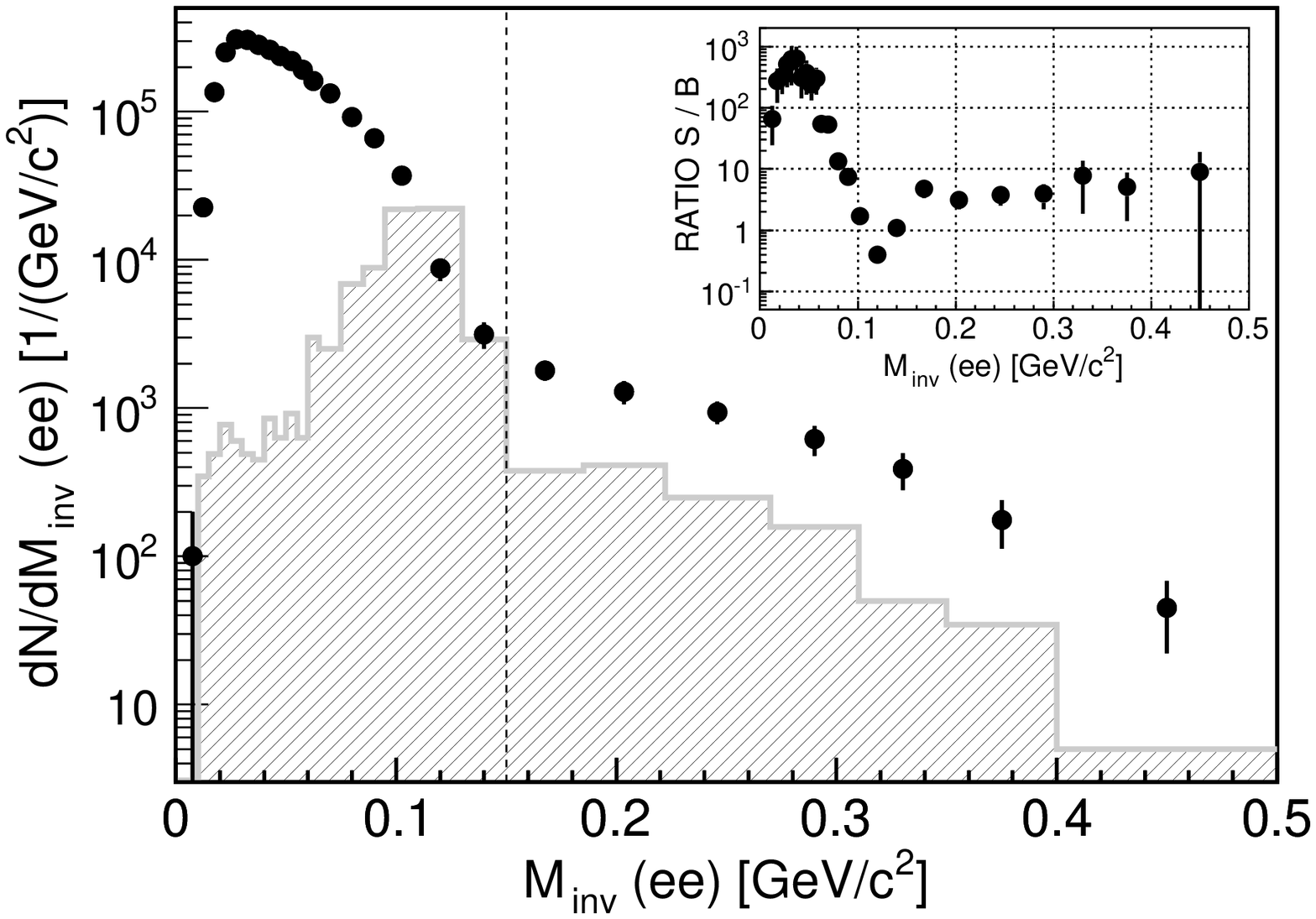}
    \includegraphics[width=0.49\textwidth]{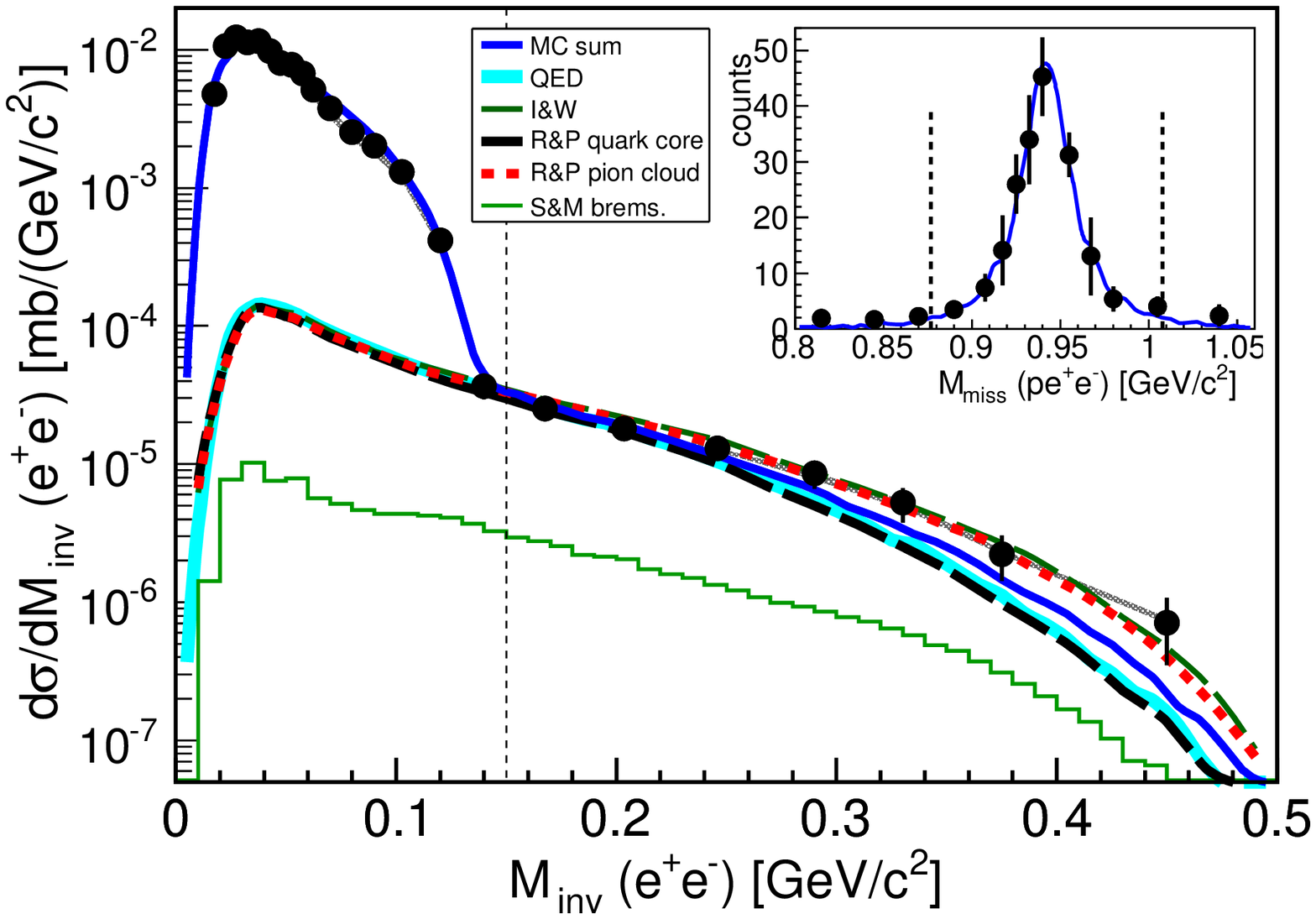}  
  %\vspace*{-0.2cm}
  \caption {(Color online) $ppe^{+}e^{-}$ final state. Left: invariant mass spectrum of $e^{+}e^{-}$ signal pairs (uncorrected data points, number of counts is per GeV/c$^2$) within the $pe^{+}e^{-}$ missing mass selection window. The gray hatched area represents the combinatorial background (CB). Inset: signal to background ratio. Right: $e^{+}e^{-}$ invariant mass within the HADES acceptance. Experimental data (black dots) are corrected for the detection and reconstruction inefficiencies (the gray thin band indicates the uncertainty of corrections, for more details see text). Blue curve represents the sum of the following contributions: $\pi^{0}$ Dalitz decay, $\Delta$ Dalitz decay according to \cite{Ramalho2016} and bremsstrahlung according to \cite{Shyam2010}. The cyan curve represents the $\Delta$ Dalitz contribution in a description with a point-like $\gamma^{*}NR$ coupling ("QED-model") \cite{Zetenyi2003,Dohrmann2010}. The two-component Iachello-Wan model \cite{IachelloWan2004,IachelloWan2005,Wan2007}, depicted with dashed dark green curve, has the largest contribution. The two components of the Ramalho-Pe\~{n}a model \cite{Ramalho2016} are shown after scaling each of them up to the same yield as in the full model: quark core (dashed black curve) and pion cloud (dashed red curve). All model contributions are supplemented with the bremsstrahlung (shown also separately as a green histogram). Normalization error is not indicated. The vertical dashed line at 0.15 GeV/c$^2$ divides the area of $\pi^{0}$ mass and higher masses. Inset: $pe^{+}e^{-}$ missing mass for $M_{inv}^{e^+e^-} >$ 0.15 GeV/c$^2$. The blue curve shows the simulation of the $\Delta$ Dalitz scaled to the same yield. Vertical dashed lines limit the window around the mass of the missing proton. 
    }
  %\vspace*{-0.2cm}
  \label{pepem_deltamass}
\end{figure*}

Figure \ref{pepem_deltamass} (left panel) shows the $e^+e^-$ invariant mass spectrum as the number of $e^+e^-$ signal pairs per GeV/c$^2$ to account for the variable bin width used. The CB is depicted as a gray hatched area. The data are plotted for a missing mass selection 0.85 $<~M_{miss}^{pe^+e^-}~<$ 1.03 GeV/c$^2$ around the proton mass (5$\sigma$ cut, see inset in Fig. \ref{pepem_deltamass}, right panel). Due to the finite reconstruction resolution a cut to reject $\pi^{0}$ Dalitz decay has been applied at $M_{e^{+}e^{-}}$ $>$ 0.15 GeV/c$^2$ (vertical dashed line). The spectrum spans up to the mass $M_{e^+e^-}\sim$ 0.5 GeV/c$^2$ which is close to the excess energy 0.54 GeV/c$^2$ available in the $p+p$ collisions for the 1.25 GeV kinetic beam energy. The signal-to-background ratio in the area above $M_{\pi^{0}}$ reaches 7-10 (Fig. \ref{pepem_deltamass}, inset in left panel). The number of reconstructed $e^+e^-$ pairs amounts to $\sim$15500 below 0.15 GeV/c$^2$ and strongly depends on the missing mass $M_{miss}^{pe^+e^-}$ selection window. The variation of the window size shows, however, that it introduces a systematic error of less than 10\% as compared to simulation. The number of $e^+e^-$ pairs for $M_{e^{+}e^{-}}$ $>$ 0.15 GeV/c$^2$ amounts to 209 pairs only. It is not dependent on the missing mass cut unless the selection window is at least 3$\sigma$. Figure \ref{pepem_deltamass} (right panel inset) shows that the Monte Carlo simulation (blue curve) of the $\Delta$ Dalitz decay gives a very similar resolution as the experimental data reconstruction. 

Figure \ref{pepem_deltamass} (right panel) presents the invariant mass spectrum of $e^+e^-$ and Fig. \ref{pepem_invmass} displays the invariant mass spectrum of $pe^+e^-$ (equivalent to missing mass of $pp \to  pe^+e^-X$) for $M_{inv}^{e^+e^-} > $ 0.15 GeV/c$^2$, within the HADES acceptance, respectively. Both spectra are corrected for the detection and reconstruction inefficiencies. The experimental data corrected with various models span over the gray band which defines the systematic (root-mean-square) error due to the model dependent inefficiency correction (see Sec.~\ref{pepem_mc}). The correction factor for masses larger than the $\pi^{0}$ mass is essentially almost constant and amounts to about 11. The $pe^+e^-$ invariant mass (Fig. \ref{pepem_invmass}) does not display the usual $\Delta$ resonance shape with the peak at 1.232 GeV/c$^2$ mass due to the selection of events with $M_{inv}^{e^+e^-} > $ 0.15 GeV/c$^2$, what naturally favours high $pe^+e^-$ masses and results in a distorted $\Delta$ spectral function. In addition, the distribution is smeared, since the proton not coming from the resonance is also included.

To justify that the data reveal the $\Delta$ resonance properties despite the unavoidable smearing due to the indiscernibility of the protons, the following distributions are studied: angular distributions of $pe^{+}e^{-}$ (missing $p$) in the c.m.s. system (Fig.~\ref{pepem_angular}, left panel) and angles between $e^+$ or $e^-$ in the $\gamma^{*}$ rest frame and the $\gamma^{*}$ itself, where dielectrons and $\gamma^{*}$ are boosted to the $\Delta$ rest frame (Fig.~\ref{pepem_angular}, right panel). This angle is measured with respect to the momentum of the $\gamma^*$ in the $\Delta$ reference frame. Both projections were corrected for the reconstruction inefficiencies and the detector acceptance, each distribution with the respective one-dimensional correction function. As above, the gray band reflects the uncertainty due to model-dependent corrections (see Sec.~\ref{pepem_mc}). Vertical black error bars reflect the statistical error only and blue horizontal bars indicate the normalization error.

\subsection{\bf{Monte Carlo simulation and results}}
\label{pepem_mc}

\begin{figure}[!htb]
  \centering
    \includegraphics[width=0.49\textwidth]{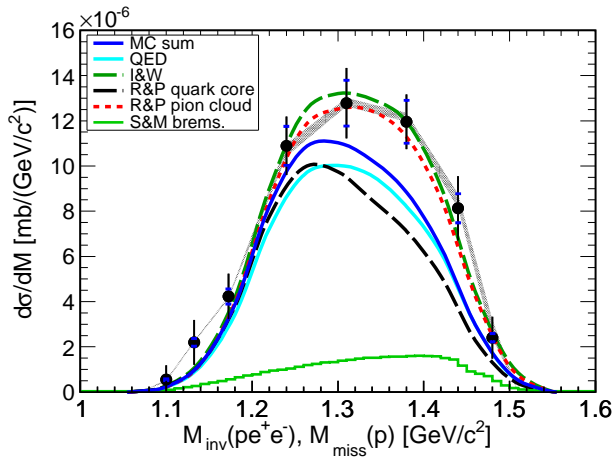}
  %\vspace*{-0.2cm}
  \caption {(Color online) $ppe^{+}e^{-}$ final state for the $M_{inv}^{e^+e^-} > $ 0.15 GeV/c$^2$. Invariant mass of $pe^{+}e^{-}$ and missing mass of a proton within the HADES acceptance (data are corrected for the reconstruction efficiency and plotted with the variable bin width). The gray band indicates the uncertainty of model-dependent one-dimensional efficiency and acceptance corrections (for details see text). Vertical error bars represent statistical error, blue horizontal bars indicate normalization error. Monte Carlo simulations (curves): blue curve represents the sum of the $\Delta$ Dalitz decay according to \cite{Ramalho2016} and non-resonant nucleon-nucleon part of bremsstrahlung according to \cite{Shyam2010} (solid green line histogram). Color codes of the other curves are as in Fig. \ref{pepem_deltamass}.
    }
  \vspace*{-0.3cm}
  \label{pepem_invmass}
\end{figure}

To estimate the contribution of $\pi^{0}$ Dalitz decay in the $pe^+e^-$ channel, corresponding analysis cuts were applied to simulated events generated with the same model as for the $ppe^+e^-\gamma$ analysis (Sec.~\ref{ppepem_mc}). It can be observed that the $e^+e^-$ invariant mass in $\pi^{0}$ region is described very well by the Monte Carlo simulation within the HADES acceptance (Fig.~\ref{pepem_deltamass} right panel). This proves the consistency of the 3- and 4-prong analyses and the very detailed description of the $pe^+e^-$ missing mass resolution, since, as observed above, the yield in this region is strongly dependent on the missing mass cuts. 

The experimental data are confronted with two descriptions of the $\Delta$ eTFF. Firstly, a point-like $\gamma^{*}NR$ model, described in Sec.~\ref{eTTF_models}, is used ("QED model"). The second model is a two-component covariant model by Ramalho-Pe\~{n}a \cite{Ramalho2016}. In all cases, the $\Delta$ resonance parametrisation and production is taken from the PWA solution (as discussed in Sec.~\ref{ppepem_mc}) as well as the cross section $\sigma_{\Delta}$ = 4.45 $\pm$ 0.33 mb. The $\Delta$ Dalitz decay is then implemented using the differential decay width calculated as a function of the running mass of the resonance and of the $e^+e^-$ invariant mass in the description of the Krivoruchenko formula \cite{Krivoruchenko2002b,Dohrmann2010} (consistent with \cite{Zetenyi2003}).

Besides the dominant $\Delta$ resonance contribution, a non-resonant virtual photon emission is added to the description, referred to as nucleon-nucleon bremsstrahlung. As discussed in Sec.~\ref{nn_bremss}, the models provide the total $e^+e^-$ contribution based on a coherent sum of many graphs describing the $\Delta$ resonance and the nucleon-nucleon bremsstrahlung contributions. In our simulation we have used the Shyam and Mosel model, which describes better data in $pp$ and $pn$ collisions at 1.25 GeV \cite{Hades2010}. It predicts the relative contribution of the nucleon-nucleon bremsstrahlung to $\Delta$ production on a level of 9\%. It is presented as a green line histogram in Fig.~\ref{pepem_deltamass} (right panel) and also in Fig.~\ref{pepem_invmass}. The contribution of the $N(1440)$ Dalitz decay can be neglected \cite{Wolf1990}.

\begin{figure}[!htb]
  \centering
    \includegraphics[width=0.49\textwidth]{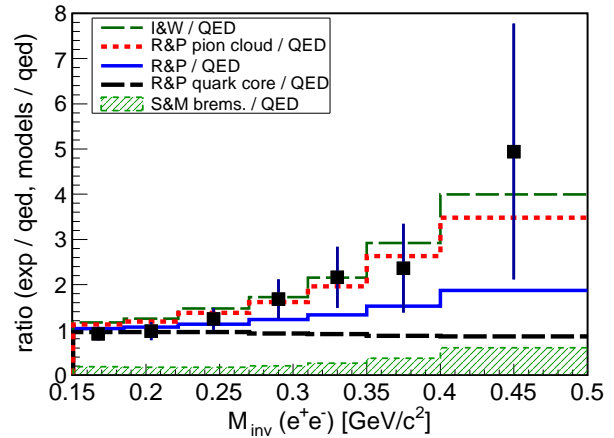}
  %\vspace*{-0.2cm}
  \caption {(Color online) $ppe^{+}e^{-}$ final state. The ratio of the experimental data (squares with error bars) to the simulated contribution of the $\Delta$ resonance with a point-like form factor ("QED model") \cite{Dohrmann2010} as a function of the invariant mass of $e^{+}e^{-}$. The experimental data are after subtraction of the simulated non-resonant nucleon-nucleon part of bremsstrahlung according to \cite{Shyam2010}. Vertical error bars represent statistical error only. Monte Carlo simulations (histograms) are also divided by the "QED model" contribution: dashed dark green line represents the Iachello-Wan model \cite{IachelloWan2004}, blue line represents the $\Delta$ Dalitz decay according to Ramalho-Pe\~{n}a model \cite{Ramalho2016}, and dotted red and black dashed lines display the pion cloud and the bare quark core components of the Ramalho-Pe\~{n}a model, respectively, after normalization to the same yield as the full model. In addition, the ratio of the bremsstrahlung to the "QED model" contribution (green hatched area) is shown as a part subtracted from the experimental data. Distributions are plotted with the same variable bin width as in Fig.~\ref{pepem_deltamass}, right panel. 
    }
  \vspace*{-0.2cm}
  \label{delta_FF}
\end{figure}

The solid cyan curve in Fig.~\ref{pepem_deltamass} (right panel) represents the simplest case: the $\Delta$ contribution with a point-like $\gamma^{*}N\Delta$ form factor \cite{Dohrmann2010} and the nucleon-nucleon bremsstrahlung \cite{Shyam2010}. The "QED model" can be considered as a lower level estimate of the $\Delta$ contribution. The blue solid curve is the sum of the full Ramalho-Pe\~{n}a model contribution \cite{Ramalho2016} and, as above, the nucleon-nucleon bremsstrahlung. The $\Delta$ decay in this model is calculated with a mass dependent eTFF with separate contributions from the quark core and the pion cloud. The presence of the form factor enhances the $e^+e^-$ yield at large invariant masses. The model describes the data just above the $\pi^0$ mass quite well but at higher $e^+e^-$ masses the data points present still an excess above the model. A possible explanation on the origin of the $e^+e^-$ excess might be drawn from the comparison of the components in the Ramalho-Pe\~{n}a model. As already mentioned above, the form factor model is composed of two ingredients. In order to do a qualitative comparison of the shape of the distribution, both components were scaled up to the same total yield in the full solid angle. The first component, the bare quark core (supplemented by bremsstrahlung), is plotted in Fig.~\ref{pepem_deltamass} (right panel) by a black dashed curve. Its distribution is similar to the "QED model" (cyan curve). This is expected, since this part of the form factor stays constant for the four-momentum transfer squared probed in our experiment. The second component, related to the pion cloud (also supplemented by bremsstrahlung), is plotted as the dotted red curve. The distribution practically describes the data points within their error bars what might indicate that this model component has a correct $q^2$ dependence and is slightly underestimated in the model. The largest contribution is provided by the Iachello-Wan model \cite{IachelloWan2004}, supplemented by the bremsstrahlung yield (dashed dark green curve). It tends to overshoot the experimental contribution at the intermediate mass 0.14 $~<~M_{inv}(e^+e^-)~<~$ 0.28 GeV/c$^2$ while giving the good description at the high mass $M_{inv}(e^+e^-)~>~$ 0.28 GeV/c$^2$.

The same model contributions are compared with the experimental data within the HADES acceptance in Fig.~\ref{pepem_invmass}, where the invariant mass of $pe^+e^-$ (or missing mass of $p$) is presented for $M_{inv}(e^+e^-)~>~$ 0.15 GeV/c$^2$. The gray band reflects again the RMS error due to the model-dependent acceptance correction. All curves are the same as in Fig.~\ref{pepem_deltamass} (right panel). As observed above for the $e^+e^-$ invariant mass, the pion cloud part of the Ramalho-Pe\~{n}a model \cite{Ramalho2016} (plus bremsstrahlung) delivers the description closest to the data. The Iachello-Wan model (plus bremsstrahlung) has a higher contribution, however within the experimental error bars.

In order to quantify the effect of the $N-\Delta$ transition form factor, the ratio of the experimental data to the simulations using the point-like form factor ("QED model") \cite{Dohrmann2010} is shown in Fig.~\ref{delta_FF} as a function of the $e^+e^-$ invariant mass. It is integrated over the $\Delta$ mass distribution as given in Fig.~\ref{pepem_invmass}. First, the simulated contribution of the non-resonant part (bremsstrahlung) \cite{Shyam2010} is subtracted from the data (it is shown as a green hatched histogram). The comparison with the models is shown with the same color code as in Fig.~\ref{pepem_deltamass} (right panel). The Ramalho-Pe\~{n}a model (solid blue) \cite{Ramalho2016} gives a good description of the data for masses $M_{inv}(e^+e^-)~<~$ 0.28 GeV/c$^2$ but then it tends to underestimate the excess. The separated pion cloud component of this model (dotted red) is the closest to the data in the whole range. The Iachello-Wan model \cite{IachelloWan2004} describes the data also well. However, the vector meson contribution in this model is not consistent with the pion electromagnetic form factor data. The differences in the parametrization of the eTFF of the pion, discussed in Sec.~\ref{eTTF_models}, are smaller than the experimental uncertainty in the studied mass range. Since they increase with the invariant mass, they have a large impact for dilepton production at higher energies \cite{Ramalho2016,Hades2014}. The quark core component of the Ramalho-Pe\~{n}a model (dashed black) is very close to the point-like contribution, as expected (see Sec.~\ref{eTTF_models}).

\begin{figure*}
  \centering
    \includegraphics[width=0.48\textwidth]{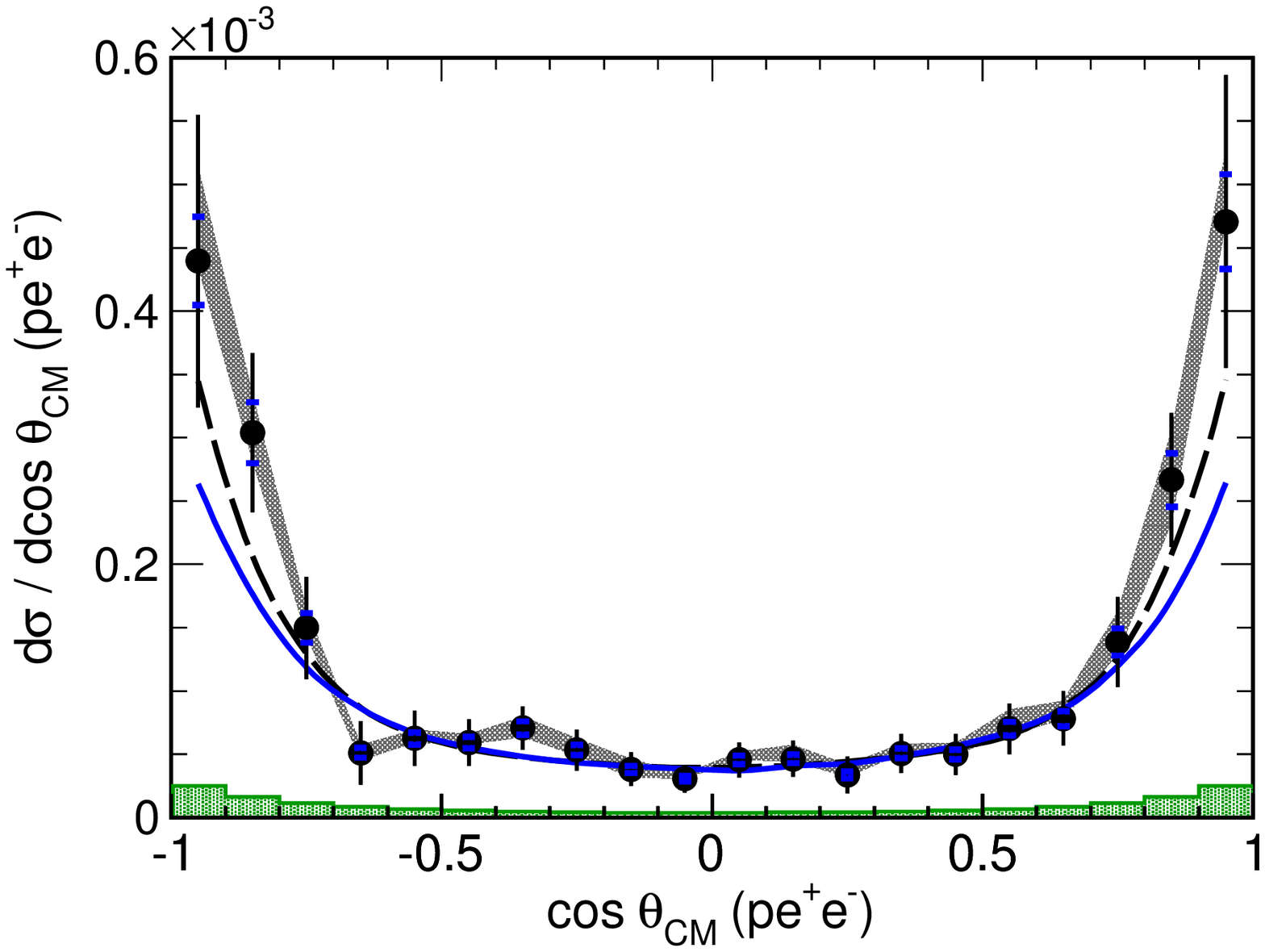}  
    \includegraphics[width=0.48\textwidth]{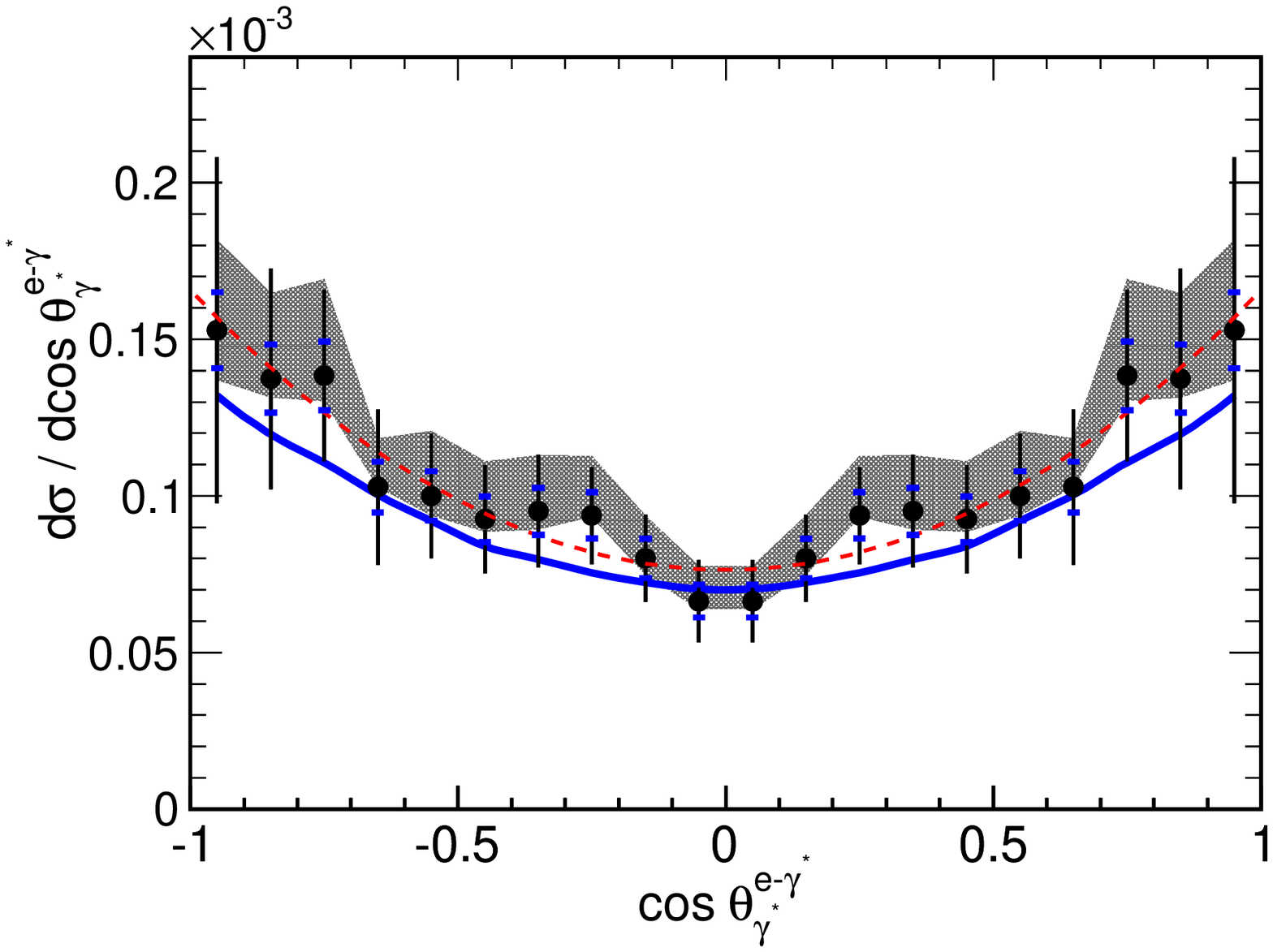}
  \vspace*{-0.1cm}
  \caption {(Color online) $ppe^{+}e^{-}$ final state for $M_{inv}^{e^+e^-}~>~$ 0.15 GeV/c$^2$. Experimental data (see symbols with error bars) are corrected for the acceptance and reconstruction inefficiencies. Both distributions are after subtraction of the simulated bremsstrahlung contribution. Left: angular distribution of $pe^+e^-$ (or missing $p$) in the c.m.s. (black dots), green shaded area at the bottom represents bremsstrahlung (see text for details). Right: $e^+$ and $e^-$ angle along the $\gamma^{*}$ direction (upper index) in the $\gamma^{*}$ rest frame (lower index). Red dashed curve is a fit $\sim 1 + B\cos^2\theta$, where $B$ = 1.17 $\pm$ 0.34. The gray band indicates the uncertainty introduced by the model dependent correction in both cases. Vertical error bars represent statistical error, blue horizontal bars indicate the normalization error. The blue curve in both cases denotes simulation results assuming $\Delta$ Dalitz decay according to model of \cite{Ramalho2016}. The black dashed curve (left) represents the $\Delta$ production from hadronic channel in the PWA description \cite{Przygoda2015}, renormalized to the same yield as the data points in the angular range -0.8 $~< \cos\theta_{CM}(pe^+e^-)~< $ +0.8.
    }
  \vspace*{-0.3cm}
  \label{pepem_angular}
\end{figure*}

An important observable describing the resonance is the production angle of the resonance system which is found to be very anisotropic in the c.m.s., both experimentally and in various model descriptions, i.e. one-pion exchange models \cite{Teis1997,Dmitriev1986} or PWA \cite{Ermakov2011,Przygoda2015,Ermakov2014}. The strong forward/backward peaking reflects the peripheral character of the $\Delta$ resonance excitation. Figure~\ref{pepem_angular} (left panel) presents the angular distribution of $pe^+e^-$ or missing $p$ in the c.m.s. First, the simulated bremsstrahlung contribution with the angular distribution modelled in line with the $\Delta$, depicted as the green shaded histogram at the bottom, was subtracted from the data points. The data are compared to the simulation using the Ramalho-Pe\~{n}a model \cite{Ramalho2016} (blue curve). The $pe^+e^-$ distribution is affected by the dependence of the angular distribution on the $\Delta$ mass. Indeed, the Ramalho-Pe\~{n}a model enhances the weight of heavier $\Delta$s which are produced with a flatter angular distribution. The predicted yield from this model is not sufficient in the very forward/backward parts of the angular distribution. This is consistent with the observation in the hadronic channel $pp \to p\Delta^+ \to pp\pi^0$ \cite{Przygoda2015}, where a similar underestimation of the proton c.m.s. angular distribution was observed at forward/backward angles (dashed black curve).  

Yet another important observable is the $e^+$ or $e^-$ angular distribution in the $\gamma^{*}$ rest frame from the $\Delta \to \gamma e^+e^-$ decay. According to calculations \cite{Bratkovskaya1995} it should also obey approximately a $1 + B\cos^{2}\theta$ dependence with the anisotropy factor $B$ = 1 if the contributions of longitudinal photons is negligible. Indeed, this seems to be the case in Fig.~\ref{pepem_angular}, right panel, where the experimental data were fitted (red dashed curve) resulting in $B$ = 1.17 $\pm$ 0.34 (the fitting error includes statistical error only). The blue curve represents the Monte Carlo simulation (as in the models discussed). The subtraction of the bremsstrahlung contribution modelled in the Monte Carlo simulation with a homogeneous distribution does not influence the fit result. Both angular distributions confirm the identification of the $\Delta$ resonance.

\section{\bf{$\Delta$(1232) Dalitz decay branching ratio}}
\label{delta_br}
The identification of the $\Delta$(1232) resonance in the Dalitz decay channel allows for the experimental determination of the branching ratio. The calculation is based on the yield measured as a function of the $pe^+e^-$ angle (Fig.~\ref{pepem_angular}, left panel) and is limited to the range -0.8 $~< \cos\theta_{CM}(pe^+e^-)~< $ +0.8, where both the hadronic and dielectron channels agree very well and systematic errors due to acceptance correction are lowest. One difficulty is related to the fact that the experimental $\Delta$ Dalitz decay yield is measured for $e^+e^-$ invariant masses $M_{inv}^{e^+e^-}~>~$ 0.15 GeV/c$^2$, which favours high $\Delta$ masses, as observed in Fig.~\ref{pepem_invmass}. In addition, due to the indiscernibility of two protons, the mass of the resonance cannot be reconstructed in a unique way. Nevertheless, simulations can be used to deduce the branching ratio at the pole from the measured Dalitz decay yield. In this purpose, we have used simulations based on the constituent covariant quark model \cite{Ramalho2016} and QED model \cite{Zetenyi2003,Dohrmann2010}, which describe the shapes of the experimental distributions very well. In addition, both simulations are based on the $\Delta$ production amplitudes deduced from the $\Delta$ pionic decay channels via the PWA. Thus, they can be safely used to extrapolate the Dalitz decay yield to the whole phase space. Both models provide a branching ratio value at the pole mass 1.232 GeV/c$^{2}$, BR($\Delta \rightarrow pe^+e^-$) = 4.2$\times$10$^{-5}$.

The procedure for deducing the branching ratio is hence enforced in the following steps:
\begin{itemize}
\item The experimental yield $N_{exp}$ (after the bremsstrahlung subtraction) in the range of -0.8 $<~\cos\theta~< $ +0.8 is calculated
\item Similarly, the integrated yield of simulated events $N_{model}$ (QED model as well as Ramalho-Pe\~{n}a model) is extracted
\item The branching ratio at the pole position is calculated by scaling the known BR of the models by the ratio of the integrated experimental and the model yields:
\begin{equation}
BR_{exp}(\Delta \rightarrow pe^+e^-) = 4.2\times 10^{-5}~\frac{N_{exp}}{N_{model}}.
\label{e_BR}
\end{equation}
\end{itemize}

The obtained $\Delta$ Dalitz branching ratio at the pole position is equal to 4.19$~\times $ 10$^{-5}$ when extrapolated with the help of the Ramalho-Pe\~{n}a model \cite{Ramalho2016}, which is taken as the reference, since it describes the data better. The branching ratio deduced with the QED model differs by 6\%. The estimate of the branching ratio also depends on the cross section for the $\Delta$ production drawn from the PWA solution wihich is affected by the error of 7.4\% ($\sigma_{\Delta} = 4.45 \pm 0.33$). Both contributions are included in the systematic error due to model uncertainty which amounts in total to 10\%. Note that we excluded from the systematic error of the PWA solution the error due to the normalisation of the data, since the same error affects the dielectron yield. Systematic errors related to the data reconstruction are similar as presented in Sec.~\ref{ppepem_mc}. Contributions to the systematic error, studied carefully by means of a Monte Carlo simulations, are due to the absolute time reconstruction, particle identification, rejection of $\gamma$ conversion, CB subtraction, missing mass window cut, efficiency and acceptance correction uncertainty. All errors, added quadratically, result in a total systematic error of 11\%. The statistical error amounts to 8\%. Finally, we arrive at the branching ratio BR($\Delta \to  pe^+e^-$) = (4.19 $\pm$ 0.42 model $\pm$ 0.46 syst. $\pm$ 0.34 stat.) $\times $ 10$^{-5}$.

\section{Summary and outlook}
\label{endsummary}
The $pp \to ppe^+e^-\gamma$ and $pp \to ppe^+e^-$ reactions have been studied in experiments using a proton beam with an incident energy of 1.25 GeV. The $ppe^+e^-\gamma$ channel accessible by HADES allows to study the $\pi^0$ Dalitz decay and to control in an independent way the $\Delta$ contribution. All distributions are in a perfect agreement with expectations from simulations. In particular, the angle between $e^+$ or $e^-$ and $\gamma^{*}$ in the $\gamma^{*}$ rest frame follows the $1+\cos^2\theta$ distribution predicted for the decay of pseudo-scalar mesons. Moreover, the yield is consistent with the measurements in the $pp \to pp\pi^0$ channel, where the $\pi^0$ was identified by the missing mass technique \cite{Przygoda2015}.

These results are used for the analysis of the $pp \to ppe^+e^-$ channel which allows to extract, for the first time, the branching ratio of the $\Delta$ Dalitz decay (4.19 $\pm$ 0.62 syst. incl. model $\pm$ 0.34 stat.) $\times $ 10$^{-5}$. The value is found to be in agreement with estimates based on calculations, using either constant electromagnetic form factors \cite{Zetenyi2003,Dohrmann2010} or a quark constituent model \cite{Ramalho2016}. 

Our work constitutes the first detailed study of a time-like electromagnetic baryon transition using the Dalitz decay process. It paves the way to the study of higher resonances, where larger four-momentum transfer can be reached and, therefore, a larger sensitivity to electromagnetic form factors could be observed. This can be achieved with HADES and the pion beam at GSI \cite{Hades2016b}. Such studies constitute an indispensable complement to measurements of space-like transitions using meson electro-production experiments. The global description of baryon transitions in both space-like and time-like regions is indeed an important challenge for the understanding of the strong force in the different energy regimes. The time-like region is particularly well suited to understand the role of vector mesons in the electromagnetic couplings. 

\begin{acknowledgments}
The \hades collaboration gratefully acknowledges the support by the grants LIP Coimbra, Coimbra (Portugal) PTDC/FIS/113339/2009, UJ Kraków (Poland) NCN 2013/10/M/ST2/00042, TU M\"unchen, Garching (Germany) MLL M\"unchen: DFG EClust 153, VH-NG-330 BMBF 06MT9156 TP5 GSI TMKrue 1012 NPI AS CR, Rez, Rez (Czech Republic) GACR 13-06759S, NPI AS CR, Rez, USC - S. de Compostela, Santiago de Compostela (Spain) CPAN:CSD2007-00042, Goethe University, Frankfurt (Germany): HA216/EMMI HIC for FAIR (LOEWE) BMBF:06FY9100I GSI F\&E, IN2P3/CNRS (France). The work of A.V. Sarantsev is supported by the RSF grant 16-12-10267.  
\end{acknowledgments}

%\footnotesize
%==============================================================================

%==============================================================================
\end{document}